\documentclass[sigconf,nonacm]{acmart}

\usepackage{multirow}
\usepackage{subcaption}
\newcommand{\del}[1]{\iffalse\st{#1}\fi}

\AtBeginDocument{%
  }

\copyrightyear{2026}
\acmYear{2026}
\setcopyright{cc}
\setcctype{by}
\acmConference[IUI '26]{31st International Conference on Intelligent User Interfaces}{March 23--26, 2026}{Paphos, Cyprus}
\acmBooktitle{31st International Conference on Intelligent User Interfaces (IUI '26), March 23--26, 2026, Paphos, Cyprus}
\acmPrice{}
\acmDOI{10.1145/3742413.3789224}
\acmISBN{979-8-4007-1984-4/2026/03}

\begin{document}

\title{How Users Perceive Mixed-Initiative AI: Attitudes Toward Assistance in Problem Solving}


\author{Yunhao Luo}
\email{yunhaoluo@ucsb.edu}
\orcid{0009-0004-6219-8021}
\affiliation{%
  \department{Computer Science}
  \institution{University of California}
  \city{Santa Barbara}
  \state{CA}
  \country{USA}
}

\author{Arthur Caetano}
\email{caetano@ucsb.edu}
\orcid{0000-0003-0207-5471}
\affiliation{%
  \department{Computer Science}
  \institution{University of California}
  \city{Santa Barbara}
  \state{CA}
  \country{USA}
}

\author{Avinash Ajit Nargund}
\email{anargund@ucsb.edu}
\orcid{0009-0005-4224-3240}
\affiliation{%
  \department{Electrical and Computer Engineering}
  \institution{University of California}
  \city{Santa Barbara}
  \state{CA}
  \country{USA}
}

\author{Tobias Höllerer}
\email{holl@cs.ucsb.edu}
\orcid{0000-0002-6240-0291}
\affiliation{%
  \department{Computer Science}
  \institution{University of California}
  \city{Santa Barbara}
  \state{CA}
  \country{USA}
}

\author{Misha Sra}
\email{sra@cs.ucsb.edu}
\orcid{0000-0001-8154-8518}
\affiliation{%
  \department{Computer Science}
  \institution{University of California}
  \city{Santa Barbara}
  \state{CA}
  \country{USA}
}

\renewcommand{\shortauthors}{Luo et al.}

\begin{abstract}
    In mixed-initiative systems, the mode of AI assistance delivery can be as consequential as the assistance itself. We investigated two assistance delivery modes: on-demand help (users request via Button) and pre-scheduled help (assistance delivered at user-selected intervals, with user actions resetting the Timer). To evaluate these modes, we selected Rush Hour puzzles as the human–AI collaborative task because they capture elements of real-world problem solving such as analysis, resource management, and decision-making under constraints. To enhance ecological validity, we imposed monetary costs for both time and AI assistance, simulating scenarios where people must balance implicit or explicit trade-offs such as time pressure, financial limitations, or opportunity costs. Although task performance was comparable across modes, participants who used the pre-scheduled (Timer) mode reported more positive perceptions of the AI, even when their ending budget was low. This suggests that assistance delivery mode can shape user experience independent of task outcomes, indicating that human-AI systems may need to consider how AI assistance is delivered alongside improving task performance.

\end{abstract}




\maketitle

\section{Introduction}

A central challenge in mixed-initiative systems, where humans and intelligent systems jointly control task execution, lies in determining the appropriate timing of assistance under uncertainty about user intentions and attentional state \cite{horvitz1999principles}. This timing problem manifests across domains. Navigation systems may reroute drivers mid-journey, misinterpreting a scenic detour as inefficiency. Educational platforms risk encouraging overreliance by offering hints that solve the problem too early or too directly~\cite{koedinger2007exploring}. Healthcare applications can send reminders at inopportune times, such as during work or late at night, reducing the likelihood that users will act on them. In each case, poorly timed assistance can undermine user agency, disrupt cognitive processes, or miss moments when help is genuinely needed~\cite{nourani2021anchoring, wen2022sense}.

Horvitz's framework on mixed-initiative interfaces proposed balancing user-initiated and system-initiated interactions, arguing for approaches that dynamically allocate initiative \cite{horvitz1999principles}. Building on this framework, contemporary systems typically implement two assistance modes: on-demand (reactive) help, where users maintain full control by explicitly requesting assistance, and proactive (anticipatory) help, where systems anticipate needs based on contextual cues \cite{newman2022helping, nair2025solicit}. On-demand assistance preserves user autonomy but requires users to interrupt their workflow and recognize when they need help. Proactive assistance can improve timeliness and reduce user burden, but faces a fundamental challenge: AI systems struggle to accurately infer user intent and attentional state \cite{bodonhelyi2024user}. This can lead to intrusions at inopportune moments that diminish a user's sense of control \cite{le2018agency, berberian2012automation}.

To address the limitations of fully automated proactive timing, many real-world systems now adopt hybrid approaches that delegate timing decisions to the AI while giving users control over the parameters that govern when and how assistance is provided. These hybrids take two primary forms. Scheduled assistance triggers at fixed, user-defined intervals, such as medication reminders that alert users every few hours~\cite{rooney2009medication, huang2024construction, paul2024smart} or productivity tools like OpenAI Tasks in ChatGPT~\footnote{\url{https://help.openai.com/en/articles/10291617-tasks-in-chatgpt}} that perform recurring actions or send reminders at preset times. Inactivity-triggered assistance or idle-time assistance, on the other hand, responds to behavioral cues, providing help when the system detects prolonged pauses or lack of engagement. For example, tutoring systems can surface hints after periods of inactivity~\cite{fossati2010generating}, or safety features like Apple's Check In~\footnote{\url{https://support.apple.com/guide/personal-safety/use-check-in-for-messages-ips56b5bc469/web}} escalate alerts when no response is detected. While both approaches aim to balance timeliness with user control, inactivity-triggered modes are particularly promising because they adapt to user behavior in real-time rather than operating on fixed schedules. Yet despite their prevalence in practice, we lack an understanding of whether and how inactivity-triggered assistance impacts user experience compared to pure on-demand help. 

Understanding these dynamics is critical for designing adaptive systems that are both effective and respectful of user autonomy, but it is complicated by two factors. First, in real-world contexts, \textit{when} help is offered is often entangled with \textit{what} form of help is provided, making it difficult to isolate the effects of timing mechanisms \cite{vereschak2021evaluate}. Second, most empirical studies have examined one-shot interactions where assistance is provided once, rather than sequential problem-solving scenarios where users engage with AI assistance repeatedly and develop adaptive strategies \cite{vantrepotte2022leveraging}. 

To explore these gaps, we conducted a controlled study examining how different assistance-timing mechanisms influence perceptions of helpfulness and problem-solving outcomes under resource constraints. We used Rush Hour puzzles\footnote{\url{https://en.wikipedia.org/wiki/Rush_Hour_(puzzle)}} as our testbed for a sequential problem-solving task where players slide blocking vehicles to free a target car. Rush Hour embodies several properties that are fundamental to a broad class of real-world problem-solving scenarios, including time pressure, the need for human reasoning in multi-step planning, and opportunities to seek help. Such characteristics are common across domains in which people increasingly collaborate with AI systems. For example, software debugging often requires stepwise hypothesis formation and testing under time constraints, where developers must decide when to inspect intermediate program states, introduce diagnostic tools, or consult documentation \cite{levin2025chatdbg}. Similarly, diagnosing and repairing household appliances involves stepwise troubleshooting, where users decide when to consult external guidance, test hypotheses, or proceed with hands-on actions~\cite{castle2025elmo}. By abstracting these shared properties into a controlled puzzle environment, Rush Hour enables the timing of assistance to be manipulated independently of assistance content, providing insights that may generalize to a broad class of human-AI collaborative problem-solving scenarios.

To instantiate resource constraints relevant to real-world help-seeking, we designed an interactive, budget-constrained environment in which a decaying monetary budget represented temporal cost (1¢ per second), and each AI intervention incurred an additional 5¢ deduction to simulate the resource cost of seeking assistance. During the study, the AI determined which cars to move based on the current puzzle state, while participants controlled when and how much help to receive through two assistance mechanisms.

In the Button condition, participants initiated assistance explicitly by requesting AI help during play and specifying the desired assistance length (e.g., a 3-move or 5-move sequence). In the Timer condition, participants initiated assistance implicitly by configuring inactivity thresholds, such that AI help was triggered automatically after a user-defined period of inactivity (e.g., ``provide 5 moves if I’m idle for 10 seconds''). Each of these conditions instantiates a canonical form of initiative allocation: user initiative and system initiative. On-demand assistance via a button operationalizes user initiative, in which users explicitly decide when to invoke automation and specify the scope of assistance, consistent with frameworks that emphasize manual invocation and direct control~\cite{parasuraman2000model, bradshaw2003dimensions}. Timer-based assistance operationalizes a constrained form of system initiative, in which intervention is triggered automatically based on behavioral cues and bounded by user-defined parameters. This design aligns with hybrid and adjustable-autonomy approaches that shift initiative to the system while preserving user oversight~\cite{bradshaw2003dimensions}. We selected these two modes because they are theoretically grounded \cite{parasuraman2000model, bradshaw2003dimensions} and widely adopted in deployed systems such as ChatGPT~\footnote{\url{https://chat.openai.com}} and Apple's Check In~\footnote{\url{https://support.apple.com/guide/personal-safety/use-check-in-for-messages-ips56b5bc469/web}}. While prior work has characterized broad paradigms of human–AI interaction (e.g., dialogic, continuous, and proactive systems)~\cite{van2021human}, our work focuses on a fundamental design dimension within these paradigms, namely how initiative over assistance onset is allocated between the user and the system. We compare the two modes under identical budget constraints to examine how different initiative allocations shape user experience, perceptions of AI helpfulness, and problem-solving performance.

Our findings ($N = 66$) reveal that assistance delivery mode shapes user experience independent of task outcomes, suggesting that how AI help is provided may matter as much as what help is provided. This has implications for both the design and evaluation of collaborative AI systems. From a design perspective, interaction modality including timing, initiative, and control mechanisms should be treated as a first-class concern alongside capability improvements. From an evaluation perspective, performance metrics alone may provide an incomplete picture, and assessments should incorporate experiential measures that capture how different interaction paradigms affect user satisfaction and sense of agency.

This work makes three contributions:
\begin{itemize}
    \item We introduce an experimental framework that decouples when help is provided from what help is provided, allowing the timing of assistance to be studied independently of content.
    \item The work provides empirical evidence that user experience, encompassing perceived autonomy, AI competence, and helpfulness, is as critical as task performance in shaping preferences for AI assistance.
    \item Our findings show that inactivity-triggered assistance is a viable design option for collaborative AI in multistep reasoning and problem-solving tasks. This finding reveals broader design implications for how timing, control, and system initiative should be balanced in human-AI collaboration.
\end{itemize}

\section{Related Work}

Our work builds on prior research at the intersection of mixed-initiative systems, and human agency in human-AI collaboration.

\subsection{Mixed-Initiative Interactions}

Human-AI collaboration has long been studied as a means to leverage the complementary strengths of humans and AI, in some cases leading to improved outcomes compared to either working alone \cite{vaccaro2024combinations}. A key paradigm in human-AI interaction is mixed-initiative interaction, where both human and AI can initiate an action, with control dynamically shared based on context and capability rather than being pre-assigned \cite{allen1999mixed}. In this line of work, Horvitz \cite{horvitz1999principles} proposed the notion of ``elegant coupling,'' the idea that assistance is most effective when delivered in synchrony with the user's context and attention. The LookOut project exemplifies this principle in a calendar system which uses text parsing and probabilistic inference to predict when a user may want to schedule an event~\cite{horvitz1999principles}. LookOut supports multiple interaction modes, including an automated assistance mode in which the system initiates actions based on its inferred user intentions. 

Realizing elegant coupling in practice requires addressing a crucial design question: when and how should system actions be triggered? Research on attention-sensitive alerting developed models for determining opportune moments for system intervention based on contextual cues \cite{horvitz2013attention}, drawing on foundational work in context-aware computing \cite{abowd1999towards} and the distinction between implicit and explicit interaction modes \cite{schmidt2000implicit}. However, empirical evidence suggests that users have varying preferences for automatic versus manual control, with the optimal balance depending on task characteristics and individual differences \cite{jameson2002pros}. These design considerations around triggering mechanisms have significant implications for how users experience and exercise agency in real-world AI systems.

Contemporary AI systems implement increasingly agentic capabilities, operating across broader contexts and supporting more complex workflows than early mixed-initiative systems \cite{saleema2019guidelines}. For example, GitHub Copilot~\footnote{\url{https://github.com/features/copilot}} can generate code suggestions based on the current file, cursor location, and recent edits. Beyond basic code completion, \citet{chen2025need} proposed a proactive AI coding assistant that can leverage both the developer's code and surrounding context, like chat history, to provide timely, integrable suggestions. Their user study showed that a proactive assistant can enhance both programmer productivity and user experience. In the domain of mental health support, \citet{liu2024compeer} introduced ComPeer, a proactive conversational peer-support agent that can self-initiate messages based on previous user interactions and regulate the frequency of its outreach. Their study demonstrated that this proactive design led to improved stress management and higher user satisfaction compared to a purely user-initiated conversational agent. Similarly, \citet{kuang2024enhancing} investigated the impact of proactive conversational AI in the domain of UX evaluation, examining how timing, whether suggestions appear before, during, or after usability problems, affects evaluator experiences. In a hybrid Wizard-of-Oz study, they found that while the overall analytic performance did not differ significantly across different timings, participants preferred suggestions that appeared immediately after a potential usability problem occurred.

Mixed-initiative research has long emphasized the value of dynamically negotiated control, with the timing of assistance identified as a critical design factor \cite{cila2022designing}. Prior work on attention-sensitive alerting illustrates one approach to deciding when a system should take initiative, using models of user attention and expected utility~\cite{horvitz2013attention}. Yet, less is known about how assistance timing should be structured and experienced within ongoing problem-solving tasks. Systems discussed earlier have relied on rule-based or heuristic methods specified by the designer to decide when to initiate assistance. In contrast, our work examines assistance timing within multistep problem solving by studying user-driven triggering mechanisms. By empirically comparing on-demand and inactivity-triggered assistance, our work complements prior mixed-initiative interaction research~\cite{horvitz1999principles, horvitz2013attention} with evidence on how assistance timing mechanisms shape user attitudes and perceptions of the AI.

\subsection{Agency and Control in Human-AI Collaboration}

Decisions about who holds initiative and control in human-AI interaction have long been framed as a critical design choice. Early theoretical frameworks established foundational concepts for understanding this distribution of control. \citet{parasuraman2000model} describes types and levels of human interaction with automation, demonstrating how sensing, deciding, and acting can be allocated across human and machine agents. Building on this foundation, research on adjustable autonomy systems identified key dimensions along which control can be dynamically negotiated between humans and autonomous agents \cite{bradshaw2003dimensions}. These theoretical insights have been translated into concrete design guidelines for human-AI interaction, including principles such as efficient invocation and dismissal, as well as mechanisms for global control that allow users to customize what the AI monitors and how it behaves \cite{saleema2019guidelines}.

Empirical studies demonstrate how these dynamics unfold in practice. A study by \citet{kang2022ai} with TikTok users found that users both welcome AI-driven personalization and intentionally influence AI algorithms to make them better cater to their needs. This bidirectional agency between humans and AI shapes patterns of user engagement and demonstrates how platform-mediated co-agency can transform user experiences. However, sustaining human agency requires more than bidirectional influence. \citet{fanni2023enhancing} argue that contestability and redress mechanisms, which allow users to challenge or reverse AI decisions, are essential to safeguarding autonomy.

As full AI automation remains limited, recent work has shifted toward characterizing human–AI collaboration. One framework proposes a unified model organized around agency, interaction, and adaptation, developed through literature synthesis and expert interviews to describe and analyze complex human–AI systems \cite{holter2024deconstructing}. However, key design questions remain underexplored, particularly regarding the temporal dimension of AI assistance. Building on this prior work, our study investigates how these dynamics play out in a collaborative problem-solving context where humans and AI jointly contribute to outcomes. We specifically decouple the timing of assistance from the amount and content of assistance, addressing a recurring challenge in mixed-initiative research where these dimensions are often conflated. Isolating timing effects enables us to derive clearer insights for informing the design of human-AI collaboration in problem-solving.

\begin{figure*}[ht]
    \centering
    \includegraphics[width=1\linewidth]{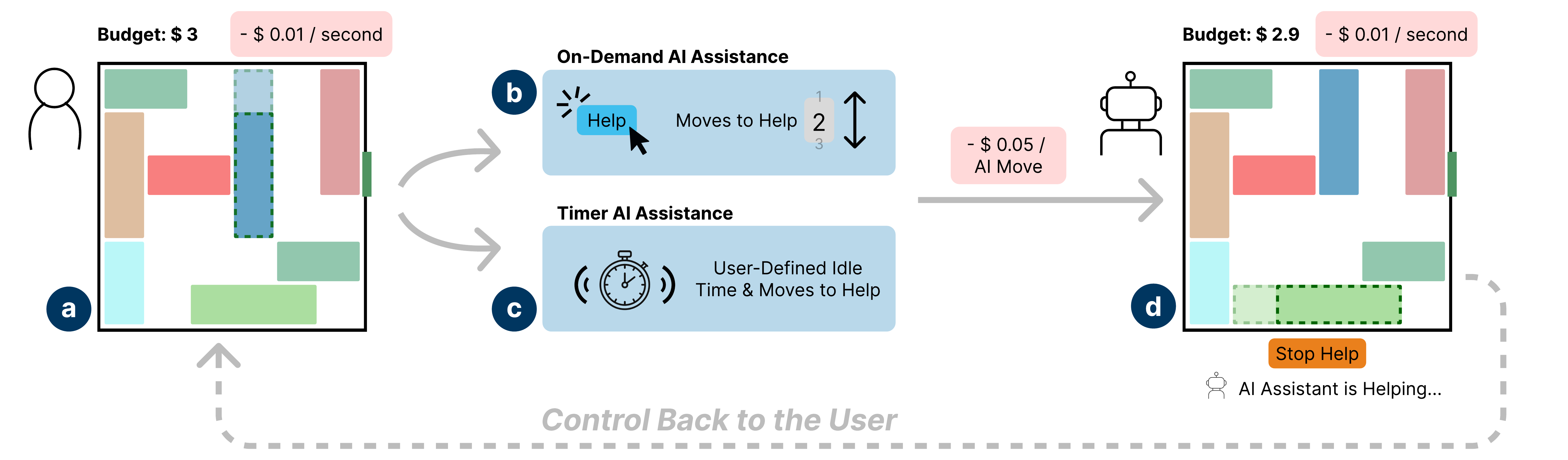}
    \caption{Study System Design. (a) Users can use their mouse to select a car (any colored rectangle), which will be highlighted with a dashed outline, and use the keyboard arrow keys to move the selected car. The two modes of AI assistance are: (b) \textit{Button mode}: Users may request help at any time during puzzle solving by clicking the Help Button and specifying the number of moves they want AI to help with. The chosen number of moves is preserved for future requests but can be adjusted at any time. (c) \textit{Timer mode}: Before starting a puzzle, users define an idle-time threshold and the number of moves they want AI to help with. If the user remains inactive (neither selecting nor moving a car) past the idle-time threshold, AI assistance is triggered. If the user selects or moves a puzzle piece before the idle-time threshold is reached, the Timer resets. These parameters cannot be changed during puzzle solving. (d) When AI assistance is triggered, the AI takes control of the board and performs the requested number of moves, with each move associated with a cost of $\$0.05$, deducted from the budget upon activation, in both conditions. While the AI assistant is active, the user may interrupt assistance by clicking on the Stop Help Button available in both conditions. There is a time cost of $\$0.01$/second in both assistance conditions.}
    \Description{The figure presents the Study System Design for a Rush Hour puzzle game with AI assistance. Panel A illustrates user control: On the left, a user is shown interacting with Rush Hour puzzle grid consists of colored rectangular blocks. The user has a visible budget of \$3, and there is a note indicating a cost of \$0.01 per second. Panel B shows On-Demand AI Assistance: A blue panel shows a button labeled "Help." The user can click this button and specify a number of "Moves to Help" using arrow controls (e.g., selecting 1, 2, or 3 moves). Each AI move costs \$0.05. Panel C shows Timer-Based AI Assistance: Another blue panel shows a click icon with labels "User-Defined Idle Time \& Moves to Help." This means the user can predefine how long they can remain inactive before the AI automatically begins helping and how many moves it should perform. Panel D demonstrates what happens when AI assistance is triggered: On the right, the AI takes control of the puzzle, performing the requested moves. A "Stop Help" button is available for the user to interrupt assistance. The budget is shown decreasing (e.g., from \$2.9), reflecting the per-second and per-move costs. A dashed arrow labeled "Control Back to the User" indicates that after the assistance ends, or if the user stops it, control returns to the user.}
    \label{fig:condition-explanation}
\end{figure*}

\section{Task Design}

To systematically investigate how different modalities of AI assistance shape collaborative problem-solving, we designed a controlled study using Rush Hour puzzles, a task that captures key characteristics of collaboration while enabling precise manipulation of when and how AI provides help. 

Rush Hour puzzles offer a well-defined, goal-oriented problem space that is both accessible and cognitively demanding. The puzzle is played on a $6\times6$ grid populated with cars one square wide and two or three squares long, oriented either horizontally or vertically. Cars can move one square at a time along their orientation, but cannot overlap or move outside the grid. The player's objective is to move the blocking vehicles to create a path for the red car to exit from the right side of the grid.

Rush Hour requires sequential and spatial reasoning, providing a structured challenge that makes patterns of help-seeking observable. The task has clear, objective metrics such as minimum moves to completion and progress, which support evaluation of performance. At the same time, the rules are easy to learn and require no prior expertise, ensuring participants can quickly learn how to do the puzzles.

\subsection{AI Assistance}

We implemented AI assistance in both modes using a deterministic breadth-first search (BFS) algorithm \cite{moore1959shortest}. BFS systematically explores the puzzle space and guarantees the shortest solution (fewest moves) from the current puzzle state. This was done to ensure consistent and optimal assistance across trials, allowing observed differences to be attributed to the mode of assistance delivery.

\subsubsection{Assistance Modes}

Based on the mixed-initiative system design framework proposed by \citet{van2021human}, which characterizes human–AI interaction in terms of different allocations of initiative between the user and the system, we implemented two assistance modes as shown in Figure~\ref{fig:condition-explanation}:

\begin{itemize}
    \item Button mode. Users request help on-demand by clicking a help Button and specifying the number of moves they want AI to help with. The requested number of moves is executed immediately, and the corresponding cost is deducted from the budget.
    \item Timer mode. Before starting each puzzle, users configure an idle-time threshold and the number of moves to receive. If they remain inactive beyond the threshold, the AI automatically intervenes with the requested number of moves. If the user moves a puzzle piece before the threshold is reached, the Timer resets. These parameters cannot be adjusted during the puzzle.
\end{itemize}

These two modes operationalize user-initiated versus system-initiated support under human instruction. The Button mode gives users direct control, letting them decide both when to request help and how much to receive. This mirrors traditional help-seeking behavior, where users consciously choose when to consult an AI resource.

In the Timer mode, assistance is triggered by user-defined periods of inactivity, introducing a proactive element that shifts some initiative to the AI. For example, if a participant sets the threshold to 10 seconds, the system monitors activity and provides help automatically after 10 seconds of no puzzle moves; any user action resets the Timer. This design addresses a key challenge in proactive AI systems where determining the optimal moment for intervention, i.e., the ``Goldilocks Time Window'' \cite{fang2025goldilocks}, remains difficult for AI to assess autonomously and risks interrupting users at inopportune times. By allowing users to define the inactivity threshold that triggers assistance, the Timer mode delegates this judgment to users while maintaining system proactivity. This approach aligns with recommendations that AI systems adapt their level of proactivity to user preferences \cite{meurisch2020exploring, cila2022designing}, and reflects real-world scenarios where autonomous agents respond to situational cues within human-specified boundaries.

\subsection{Budget Structure}
\label{subsec:budget-structure}

To simulate real-world problem-solving under time pressure and to incentivize active engagement \cite{lazar2017research}, we implemented a structured cost system, an approach that has been adopted in prior research \cite{kobis2025delegation, erlei2022s}. At the start of each puzzle, participants were given \$3, displayed at the top of the puzzle board. The budget decreased according to two cost types:

\begin{itemize}
    \item \textbf{Time cost:} every elapsed second deducts \$0.01.
    \item \textbf{AI assistance cost:} each AI move deducts \$0.05. 
\end{itemize}

This created a maximum of $300$ seconds to solve a puzzle without assistance. When AI help was requested or triggered, the corresponding cost was deducted immediately, and the budget continued to decrease with the passage of time while the AI executed moves. The AI required half a second to select a car and half a second to move it. For consecutive moves of the same car, the AI only needed to select the car once. The remaining budget at puzzle completion was recorded. If the budget reached zero before the puzzle was completed, the study continued to the next puzzle. 

The budget system was refined through pilot testing to balance fairness and challenge. Participants could maximize their earnings by solving independently or using assistance strategically when stuck. Reliance on the AI alone was not sufficient, since some puzzles could not be solved in time given the AI's move speed, and puzzle difficulty and solution length were unknown in advance. This uncertainty encouraged careful trade-offs between conserving budget and leveraging AI assistance.

\section{User Study}

We conducted a controlled online user study to examine how different assistance delivery modes influence problem-solving progress, problem-solving efficiency, and perceptions of AI assistance.

In this study, we investigated two main research questions:

RQ1: How do Button and Timer modes influence problem-solving efficiency, resource usage, and progress under time and budget constraints?

RQ2: How do these assistance modes impact user perception of AI's competence, helpfulness, and collaboration?

By addressing these questions, this work aims to advance our understanding of timing control in mixed-initiative systems, specifically how users balance autonomy, efficiency, and support when assistance carries an explicit cost.

\subsection{Participants}

To determine the required sample size, an \textit{a priori} power analysis was conducted for an ANCOVA with two experimental conditions (Button and Timer conditions) and one covariate. Assuming a large effect size of $f = 0.45$, $\alpha = 0.05$, $1-\beta = 0.80$, the analysis indicated a minimum sample size of 41. To maintain equal group sizes across the two AI-assisted conditions and the control group, we recruited a total of 66 participants (22 participants per condition, 35 male, 30 female, 1 non-binary). Participants also reported their AI usage frequency across four categories: daily ($N = 25$), a few times a week ($N = 26$), a few times a month ($N = 13$), and never ($N = 2$). We recruited participants through Prolific\footnote{https://www.prolific.com} for an online experiment conducted via a custom web interface. Eligible participants were required to have an approval rate of at least 99\%, and be fluent in English, as all study instructions and pre-/post-study questionnaires were administered in English. Each participant received \$4 base compensation for the approximately 30-minute study session, plus a task-based bonus determined by the budget system described in Section~\ref{subsec:budget-structure}. Participants received a bonus equal to their average remaining budget across the seven puzzles (two baseline puzzles and five main study task puzzles).

\begin{figure*}[ht]
    \centering
    \includegraphics[width=1\linewidth]{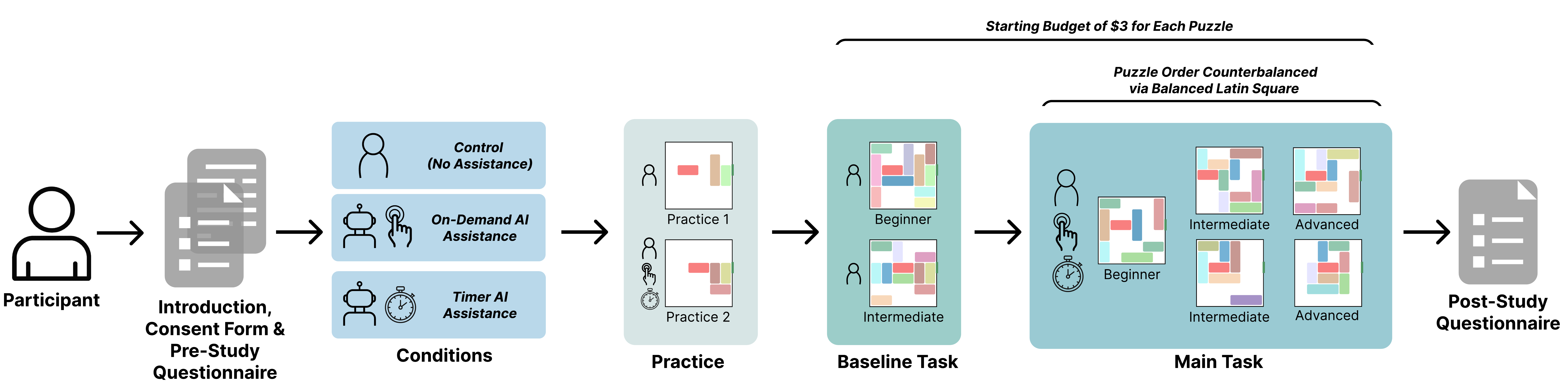}
    \caption{Study Procedure. Participants began by reading an introduction to the Rush Hour puzzle and the study objectives, followed by a pre-study questionnaire. They then completed two practice puzzles: the first without AI assistance to learn the puzzle and controls, and the second with the AI assistant from their assigned condition. After practice, all participants solved two puzzles of different difficulty levels without assistance to establish baseline puzzle-solving skills. They then proceeded to the main study phase, which involved solving five puzzles of varying difficulty, working with the AI corresponding to their assigned condition. The session concluded with a post-study questionnaire.}
    \Description{This figure presents the sequence of steps that participants followed in the study, shown as a left-to-right flow of labeled icons and panels: (1) Participant Icon: A simple figure representing the participant starting the process.
	(2) Introduction and Pre-Study Questionnaire: A clipboard-like icon represents reading an introduction to the Rush Hour puzzle and completing a consent form and questionnaire. (3) Condition Assignment: The participant is assigned to one of three AI assistance conditions: Control (No Assistance), On-Demand AI Assistance, Timer AI Assistance. These are shown as three side-by-side blue panels. (4) Practice Phase: A blue-green panel labeled Practice 1 and Practice 2 shows two puzzle boards. In the first, participants practice without AI to learn controls. In the second, they practice using the AI assigned to their condition. (5) Baseline Task: A green panel showing two puzzles labeled Beginner and Intermediate. All participants solve these without AI to establish their natural puzzle-solving ability. (6) Main Task: A larger teal panel showing five puzzles of varying difficulty: Beginner, Intermediate, and Advanced, with icons indicating that AI assistance is available (depending on condition). A note above reads "Starting Budget of \$3 for Each Puzzle" and "Puzzles Order Counterbalanced via Balanced Latin Square." (7) Post-Study Questionnaire: A clipboard icon marks the final step, where participants complete a concluding questionnaire.}
    \label{fig:study-procedure}
\end{figure*}

\subsection{Procedure}

Participants were randomly assigned to one of three conditions: (1) \textbf{Control} condition with no AI assistance; (2) \textbf{Button} condition where help could be requested via a Button; and (3) \textbf{Timer} condition where assistance was delivered after a user-defined idle threshold. See Figure~\ref{fig:study-procedure} for an illustrated study procedure. The study began with participants providing informed consent (IRB protocol \#38-25-0458) followed by an introduction screen explaining the Rush Hour puzzle rules, keyboard control method, study objectives, and the budget system. Participants then completed a pre-study questionnaire capturing demographic information, prior experience with AI tools, and the Affinity for Technology Scale (ATI) \cite{franke2019personal}. 

Next, participants completed a practice task consisting of two puzzles. For all participants, the first puzzle was completed without AI assistance to familiarize them with the interface. The second puzzle was completed under the participant's assigned condition, which was either Button, Timer, or Control with no assistance. Following practice, all participants solved two puzzles without AI assistance to establish baseline problem-solving ability. The same puzzles, with a fixed order, were used for all participants.

\begin{figure*}
    \centering
    \includegraphics[width=1\linewidth]{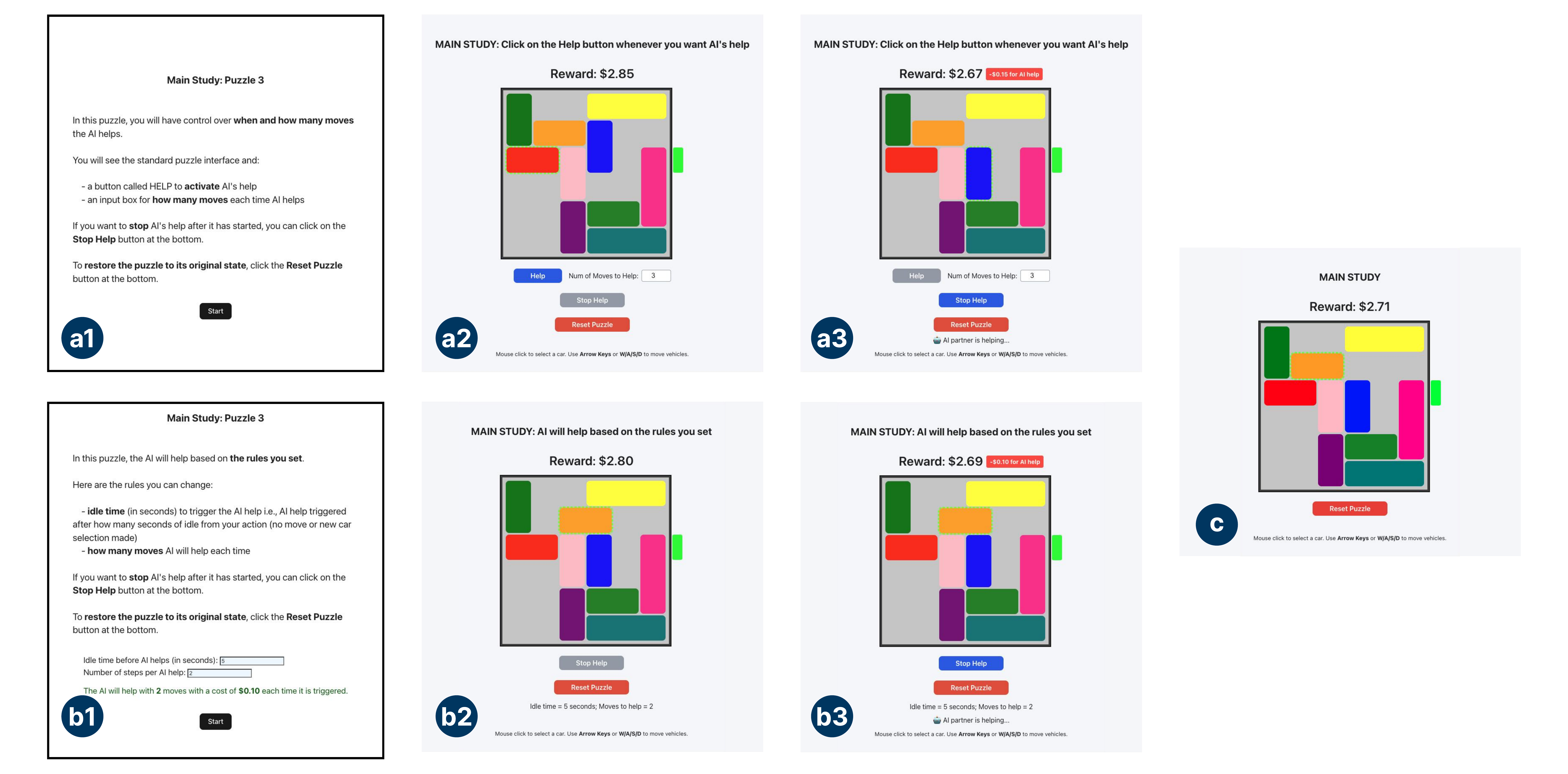}
    \caption{User interface used in the user study. (a1) Instruction panel for the Button condition. (a2) User in control in the Button condition. Users can adjust how many moves of help to receive for the next AI assistance. (a3) AI in control; users can click on the Stop Help Button to terminate assistance. (b1) Instruction panel for the Timer condition, with a configuration module below for users to set the idle-time threshold and the number of moves to help. (b2) User in control in the Timer condition. (b3) AI in control; as in the Button condition, assistance can be terminated by clicking on the Stop Help Button. (c) Interface for the control condition and baseline phase. Users may click on the ``Reset Puzzle'' button to reset the puzzle to its initial state. This reset button is also available in both the Button and Timer conditions.}
    \Description{The figure shows screenshots from the puzzle interface used in the experiment, illustrating different stages and conditions of AI assistance. The layouts are grouped by condition and labeled (a1–a3), (b1–b3), and (c). Button condition: (a1) Instruction Panel: A text-based instruction page titled "Main Study: Puzzle 3." It explains that users can request AI help by clicking a button and adjust how many moves they want assistance with. A "Start" button appears at the bottom. (a2) User in Control: A puzzle board with colored rectangular blocks is displayed. At the top, the text reads "MAIN STUDY: Click on the Help button whenever you want AI’s help" and shows a reward amount (e.g., \$2.85). Below the puzzle, there is a "Help" button and a dropdown labeled "Number of moves to help". A red "Reset Puzzle" button is also available. (a3) AI in Control: Similar layout to (a2), but the puzzle is marked as being controlled by AI. A Stop Help button appears, allowing users to terminate assistance early. Timer condition: (b1) Instruction Panel: A similar text instruction layout titled "Main Study: Puzzle 3," but includes controls to set an idle-time threshold and number of AI moves. A "Start" button appears at the bottom. (b2) User in Control: The puzzle screen shows "AI will help based on the rules you set". AI hasn’t started yet, but a Stop Help button is available once triggered. Reward amount is displayed (e.g., \$2.80). (b3) AI in Control: AI is moving the puzzle blocks automatically. As with the Button condition, a Stop Help button allows users to interrupt assistance. Reward amount updates accordingly. Control Condition: (c) No AI Assistance: Displays only the puzzle and reward at the top (e.g., \$2.71). A "Reset Puzzle" button is available, but no Help or Stop Help buttons appear.}
    \label{fig:interface}
\end{figure*}

In the main study phase, participants solved five puzzles with varying difficulty levels using the assistance mode assigned to their condition. Puzzle order was counterbalanced across participants using a balanced Latin square~\cite{bradley1958complete}. 
Before starting each puzzle, participants viewed an information panel reminding them of the study phase and how AI assistance could be requested. In the Timer assistance condition, this panel also included a configuration module for setting the idle-time threshold and the number of moves to receive help with. The puzzle was revealed only after participants dismissed the information panel, at which point the costs began to apply. Across all conditions, participants could reset a puzzle to its initial state at any time using a reset button. Upon completing all five puzzles, participants were directed to a post-study questionnaire. Screenshots of the study interface are shown in Figure~\ref{fig:interface}.

\subsection{Puzzle Selection}

All participants received the same set of puzzles, drawn from an online Rush Hour platform\footnote{http://www.mathsonline.org/game/jam.html}. The platform defines four difficulty categories: beginner, intermediate, advanced, and expert, from which we selected puzzles spanning three categories (beginner, intermediate, and advanced) to cover a range of difficulty levels while avoiding ceiling and floor effects under the fixed time constraints. The two baseline puzzles included one beginner and one intermediate puzzle, while the five main study puzzles consisted of one beginner (puzzle 1), two intermediate (puzzle 2 \& 3), and two advanced puzzles (puzzle 4 \& 5). Each puzzle has a known optimum number of moves needed to solve it; however, this information was not revealed to participants.

\subsection{Measures}

To evaluate the impact of AI assistance on problem-solving, both in comparison to human only (Control condition) and across the two AI assistance modes (Button and Timer), we collected both objective measures of performance and subjective feedback from participants on their perceptions of the assistance.

\subsubsection{Performance Measures}\label{sec:measures}

Performance was assessed across three dimensions: (1) puzzle-solving progress, (2) puzzle-solving precision, and (3) budget efficiency. To quantify these dimensions, we introduce performance metrics. For a given puzzle, we define:

\begin{itemize}
    \item $O$ = the optimum number of moves required to solve the puzzle from the starting state.
    \item $M$ = the total number of moves, including the moves made by both the human and the AI.
    \item $R_i$ = the number of moves remaining to solve the puzzle at step $i$ where $i \in [1, M]$. By definition, at the start of the puzzle or its initial state ($i=1$), $R_1 = O$.
    \item $B$ = the total amount of budget that was spent.
\end{itemize}

From these, we derive the following performance metrics:

\begin{enumerate}
    \item Puzzle-moves precision: For each puzzle, precision $P$ is defined as:
    \[
    \frac{O - min_{1\leq j \leq M}R_j}{M}
    \]
    By definition, $P=1$ when a puzzle is solved using the optimum number of moves, and $P$ decreases as additional unnecessary moves are made.
    \item Maximum progress: First, for each puzzle, progress $G$ is defined as:
    \[
    1 - \frac{min_{1\leq j \leq M}R_j}{O}
    \]
    By definition, $G = 0$ at the beginning of a puzzle and $G = 1$ upon completion. Maximum progress for each puzzle is the highest value of $G$ across all moves made.  
    \item Remaining budget: The remaining budget for each puzzle. Recorded at the time when the puzzle was solved or when the budget was fully depleted, in which case the recorded value was 0.
    \item Budget efficiency: For each puzzle, the budget efficiency is defined as:
    \[
    \frac{O - min_{1\leq j\leq M}R_j}{B}
    \]
    This metric represents the ratio between effective progress towards the solution and the budget expenditure (including time cost and AI cost). It quantifies the number of effective moves made per dollar of budget spent.
\end{enumerate}

\subsubsection{Subjective Measures}

Perceived workload was assessed using the six-item short-form NASA-TLX questionnaire \cite{hart1988development}. To capture attitudes toward the AI assistance modes, we developed a set of post-study questions, combining items adapted from established questionnaires ($N = 11$) with custom questions tailored to the puzzle-solving task ($N = 7$). The questions were adapted from: (1) two questions from the Trust Perception Scale-HRI \cite{schaefer2016measuring}, a widely used measure in human-AI collaboration research, focusing on perceived competence, reliability, and other dimensions of AI trustworthiness; (2) three questions inspired by the Technology Acceptance Model \cite{davis1989perceived}, particularly its construct of ``Perceived Usefulness,'' which evaluates the degree to which a person believes using a system will enhance their task performance; (3) two questions derived from the Teamwork Quality Scale \cite{hoegl2001teamwork}, a framework for measuring collaboration, including dimensions such as communication, coordination, and mutual support; and (4) four questions based on the perceived collaboration questions used in human-robot interaction study \cite{nikolaidis2017human}, addressing satisfaction and perceptions of control. These items allowed us to assess not only the cognitive demands of the task but also participant perceptions of the AI assistance. All questions were presented on a 7-point Likert scale (see Appendix~\ref{post-study-questionnaire} for the full list of questions). We also included custom open-ended questions to further probe participants' experience. For those in the AI-assisted conditions, these questions focused on the specific assistance delivery mode they interacted with. For participants in the control condition, the questions instead explored their general attitudes toward receiving assistance and their preferred form of AI assistance (See Appendix~\ref{post-study-open-ended} for the full list of questions).

\section{Quantitative Results}

In total, we collected data from 66 participants. Each participant completed two baseline puzzles and five main study puzzles. One puzzle from P60 (Timer, male) was discarded for analysis due to missing log data, resulting in a final dataset of 329 puzzle trials for analysis.
 
To evaluate statistical differences in performance between the three conditions, we employed generalized linear mixed models (GLMMs) implemented with the glmmTMB package in R~\cite{brooks2017glmmtmb}, selecting the family and link function based on the data distribution and model fit. For each performance metric, we specified condition and puzzle difficulty (through the optimum number of moves, $O$) as the primary predictors and used them as fixed effects. We additionally included each participant's average score on the same metric across the two Baseline puzzles as a covariate to control for individual skill levels. To account for variability across participants, participant ID was modeled as a random intercept. To maintain model interpretability, we focused on ecologically important effects only \cite{aho2013foundational}. Model convergence was checked for all fitted models, and residual diagnostics were performed using DHARMa~\cite{hartig2016dharma} to verify distribution assumptions.

To assess significance, we applied Type II Wald-square tests using the car package in R for each fixed effect~\cite{fox2018r}. If condition was a significant predictor, we conducted post-hoc pairwise comparisons using estimated marginal means with Tukey adjustment for multiple comparisons.

\begin{figure*}[ht]
    \centering
        \centering
        \includegraphics[width=\textwidth]{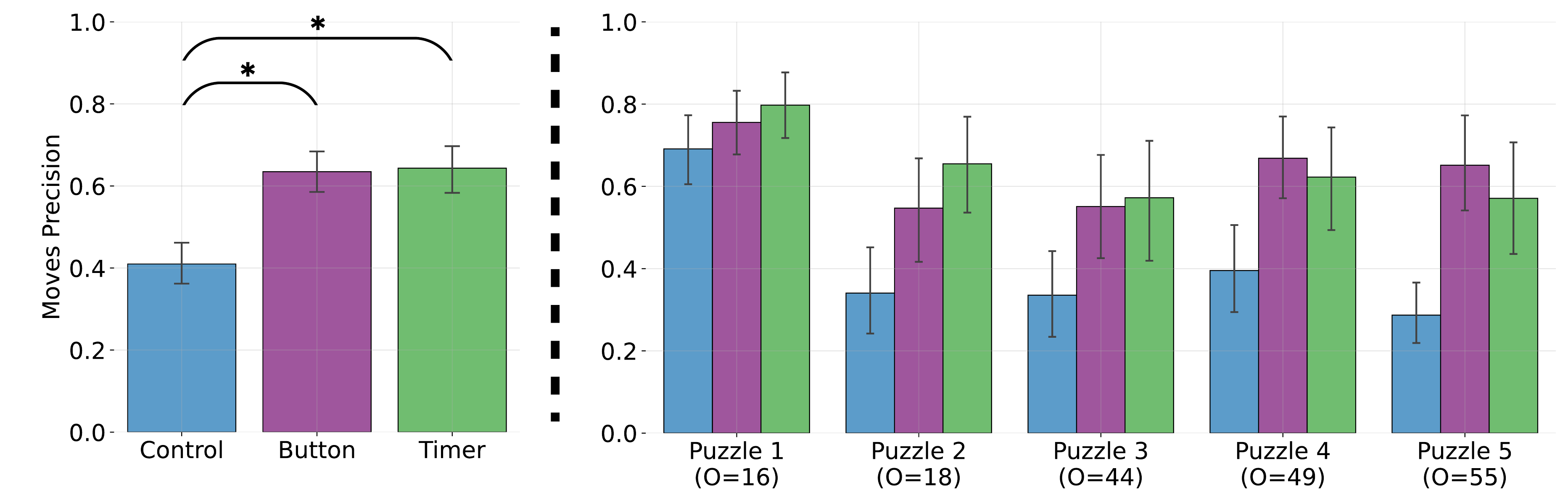}
        \caption{Moves precision for main study puzzles across conditions (left). Asterisks denote significant differences between the Button and Control conditions and between the Timer and Control conditions. Comparisons between conditions for each puzzle in the main study phase (right). $O$ denotes the optimum number of moves to solve each puzzle from its initial state. Bars represent means, and error bars denote 95\% confidence intervals; these conventions apply to all subsequent bar charts.}
        \Description{The figure consists of two grouped bar charts comparing move precision, how efficiently participants solved puzzles relative to the optimal number of moves, across three experimental conditions: Control, Button, and Timer. Left Chart: Overall precision across conditions. Three bars are shown: Control (blue): approximately 0.40 precision. Button (purple) approximately 0.65 precision. Timer (green): also around 0.65 precision. Two asterisks (*) with curved brackets indicate statistically significant differences: Button vs Control: Button users were significantly more precise than control. Timer vs Control: Timer also significantly more precise than Control. There is no significance marker between Button and Timer. Right Chart: precision by individual puzzle (five puzzles) Each group represents one puzzle, with three bars per group. Across all puzzles, Control consistently shows the lowest precision, while AI-assisted conditions (Button and Timer) produce higher precision, especially on more complex puzzles.}
        \label{fig:moves_precision}
\end{figure*}
\begin{figure*}
        \centering
        \includegraphics[width=\textwidth]{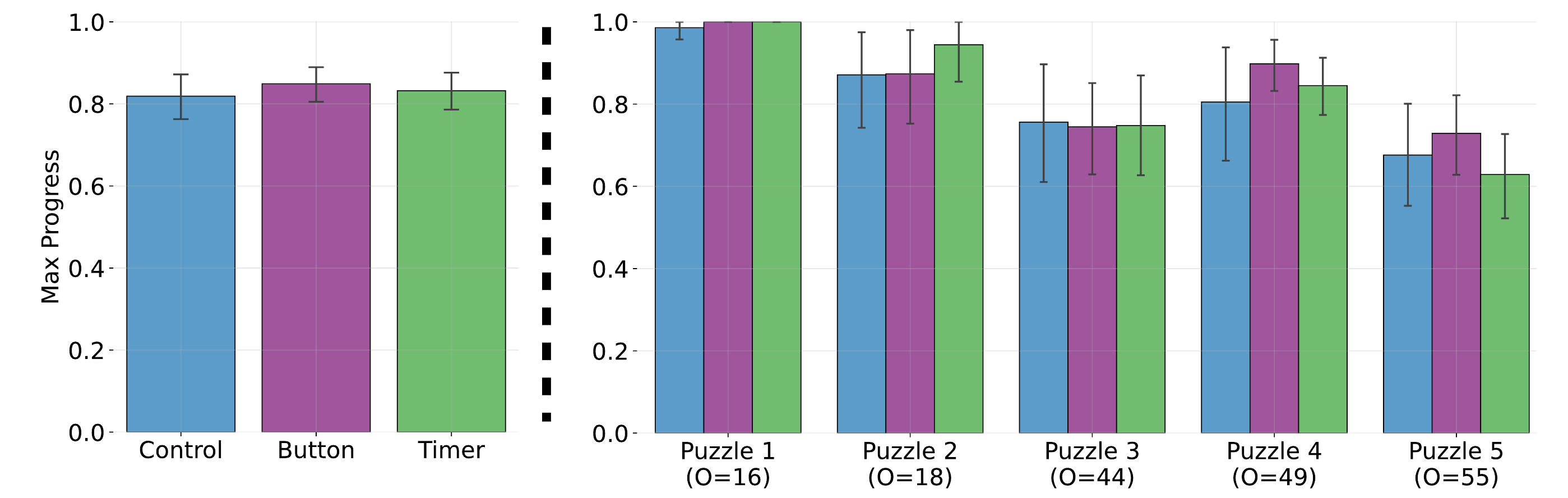}
        \caption{Maximum progress achieved for main study puzzles across conditions (left) and comparisons between conditions for each puzzle in the main study phase (right).}
        \Description{The figure consists of two grouped bar charts comparing max progress across three experimental conditions: Control, Button, and Timer. Left Chart: Overall max progress across conditions. Three bars are shown: Control (blue): approximately 0.80. Button (purple) approximately 0.82. Timer (green): around 0.81. There are no significance markers. Right Chart: max progress by individual puzzle (five puzzles) Each group represents one puzzle, with three bars per group. Across all puzzles, the three conditions show similar max progress.}
        \label{fig:max_progress}
\end{figure*}

\begin{figure*}[ht]
    \centering
    \begin{subfigure}[t]{0.45\textwidth}
        \centering
        \includegraphics[width=\textwidth]{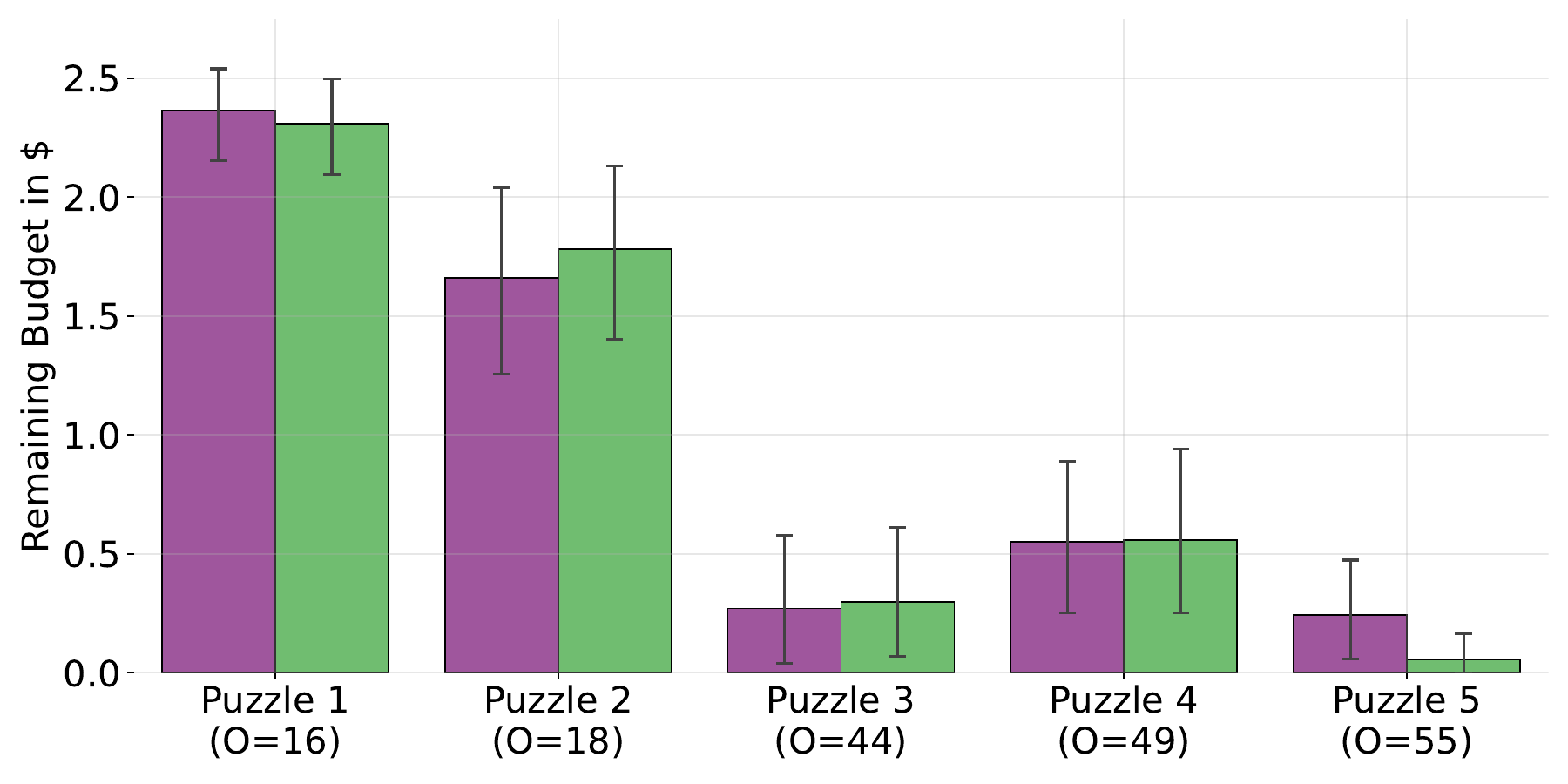}
        \caption{Remaining budget.}
        \label{fig:reward}
    \end{subfigure}
    \hfill
    \begin{subfigure}[t]{0.45\textwidth}
        \centering
        \includegraphics[width=\textwidth]{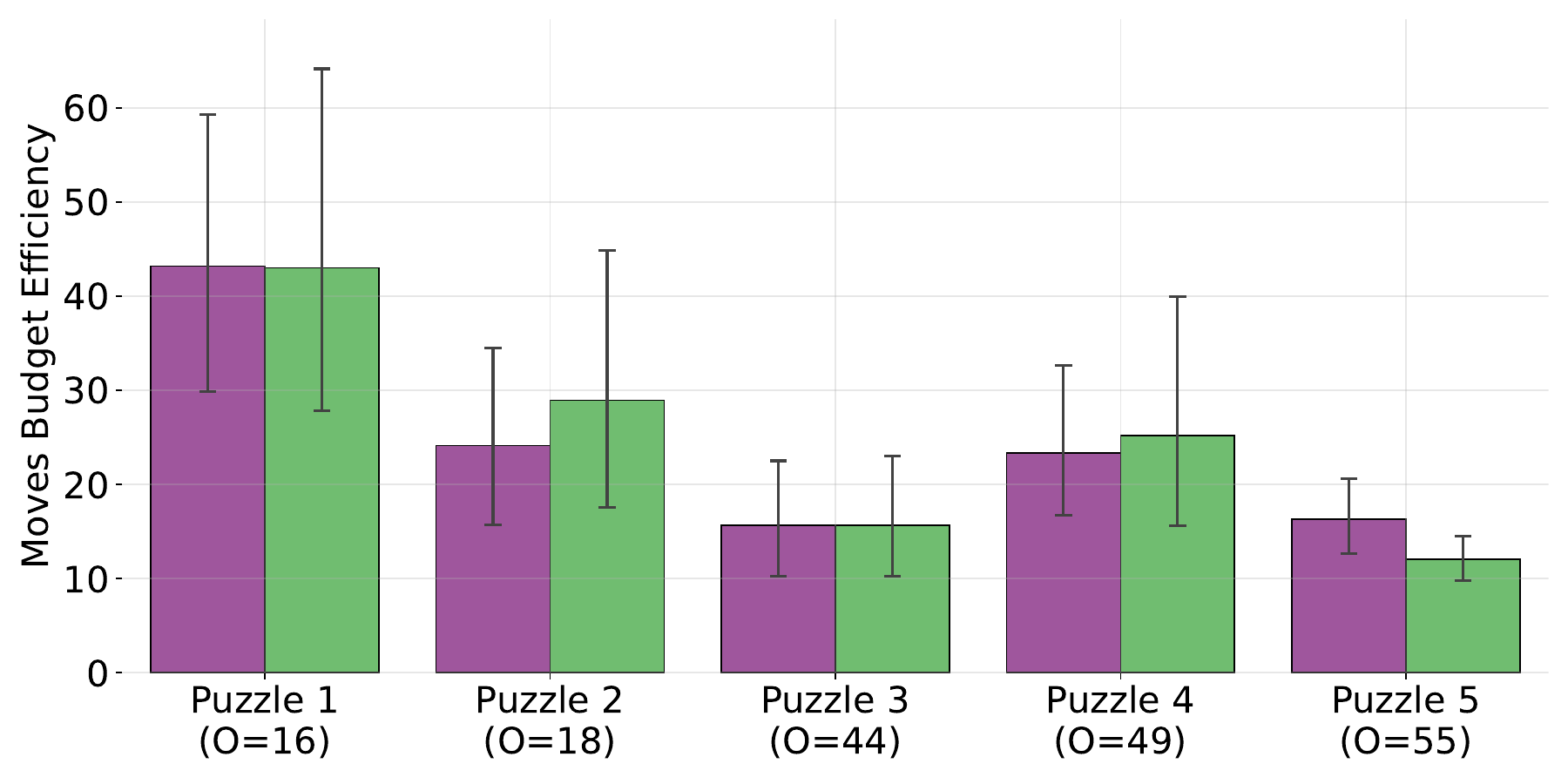}
        \caption{Budget efficiency.}
        \label{fig:budget_efficiency}
    \end{subfigure}
    \caption{Comparison of remaining budget and budget efficiency across AI-assisted conditions for main study puzzles. Purple denotes the Button condition, and green denotes the Timer condition.}
    \Description{The figure consists of two side-by-side bar charts comparing AI-assisted conditions (Button and Timer) across five puzzles. Each puzzle includes the optimal number of moves shown beneath (denoted as O). Purple bars represent the Button condition, and green bars represent the Timer condition, with error bars indicating variability. (a) Remaining Budget: This chart shows how much virtual currency (out of \3 starting budget) participants had left after completing each puzzle. Puzzle 1 (O = 16): Both Button and Timer retain high budget (~\$2.3–\$2.4), nearly identical. Puzzle 2 (O = 18): Budget decreases to around ~\$1.6 (Button) and ~\$1.7 (Timer). Puzzle 3 (O = 44): Significant drop — Button retains \$0.5, Timer slightly higher (\$0.3). Puzzle 4 (O = 49): Both remain around \$0.5. Puzzle 5 (O = 55): Lowest overall — Button retains ~\$0.3, Timer ~\$0.1.}
    \label{fig:main_budget_comparison}
\end{figure*}

\subsection{Task Performance}
We evaluated participant performance on the five main task puzzles (see Figure~\ref{fig:study-procedure}), examining differences across conditions.

\subsubsection{Moves Precision}
 Given that the puzzle-moves precision is a continuous measure bounded between 0 and 1, we employed an ordered beta regression with a logit link, a family of models designed specifically for bounded continuous or proportional data \cite{kubinec2023ordered}.

Results show significant main effects for condition ($\chi^2 = 17.654$, $p < 0.001$) and puzzle difficulty ($\chi^2 = 9.610$, $p = 0.002$). Baseline moves precision was not a significant predictor ($\chi^2 = 2.742$, $p = 0.098$). Post-hoc pairwise comparisons indicate that the control condition had a significantly lower moves precision than both the Button (odds ratio $= 0.483$, $p = 0.001$) and Timer (odds ratio = $0.463$, $p < 0.001$) conditions. No difference was observed between the Button and the Timer conditions (odds ratio $= 0.958$, $p = 0.978$). See Figure~\ref{fig:moves_precision}.

\subsubsection{Remaining Budget} 
For the remaining budget, values were converted to proportions relative to the initial $\$3$ budget, and an ordered beta regression with a logit link was applied.

Results show significant main effects of condition ($\chi^2 = 6.719$, $p = 0.035$), puzzle difficulty ($\chi^2 = 283.791$, $p < 0.001$) and baseline remaining budget ($\chi^2 = 38.032$, $p < 0.001$). Post hoc analysis shows that the Control condition had a significantly higher remaining budget than the Timer condition (odds ratio $= 1.56$, $p = 0.049$). The Control condition also shows a trend towards higher remaining budget than the Button condition, although this difference is not significant (odds ratio $= 1.50$, $p = 0.080$). No significant difference was observed between the Button and the Timer conditions (odds ratio = $1.04$, $p = 0.964$). See Figure~\ref{fig:reward}.

\subsubsection{Budget Efficiency}
For budget efficiency, a Gamma distribution with log link was applied due to its continuous and right-skewed distribution. To satisfy the Gamma distribution's requirement of strictly positive values, a small constant ($\epsilon$) was added to one data point where budget efficiency was zero (P17, Control, Puzzle 3).

The results show significant effects of puzzle difficulty ($\chi^2 = 59.680$, $p < 0.001$) and Baseline budget efficiency ($\chi^2 = 40.073$, $p < 0.001$). However, condition did not have a significant effect ($\chi^2 = 1.521$, $p = 0.468$).

\subsubsection{Maximum Progress}
Because maximum progress reflects the proportion of each puzzle solved, we initially used an ordered beta regression. However, the model failed to converge, likely due to the large number of values equal to one (fully solved puzzles), making it unsuitable for ordered beta regression. To resolve this, we transformed the variable by computing $1 - \textit{max progress}$, yielding a new measure representing the proportion of the puzzle remaining unsolved. This transformation shifted the distribution's mass from the upper to the lower bound. We then applied the Tweedie family with a log link, which is well-suited for continuous data with a spike at zero~\cite{shono2008application}.

Baseline remaining progress shows a borderline effect ($\chi^2 = 3.427$, $p = 0.064$) and puzzle difficulty shows a significant effect ($\chi^2 = 53.693$, $p < 0.001$). Condition does not have a significant effect ($\chi^2 = 0.138$, $p = 0.934$).

\subsubsection{Progress Over Time}
\begin{figure*}[htbp]
  \centering
  \captionsetup[sub]{font=small,labelfont=bf,justification=centering}

  \begin{subfigure}[t]{0.45\textwidth}
    \centering
    \includegraphics[width=\linewidth]{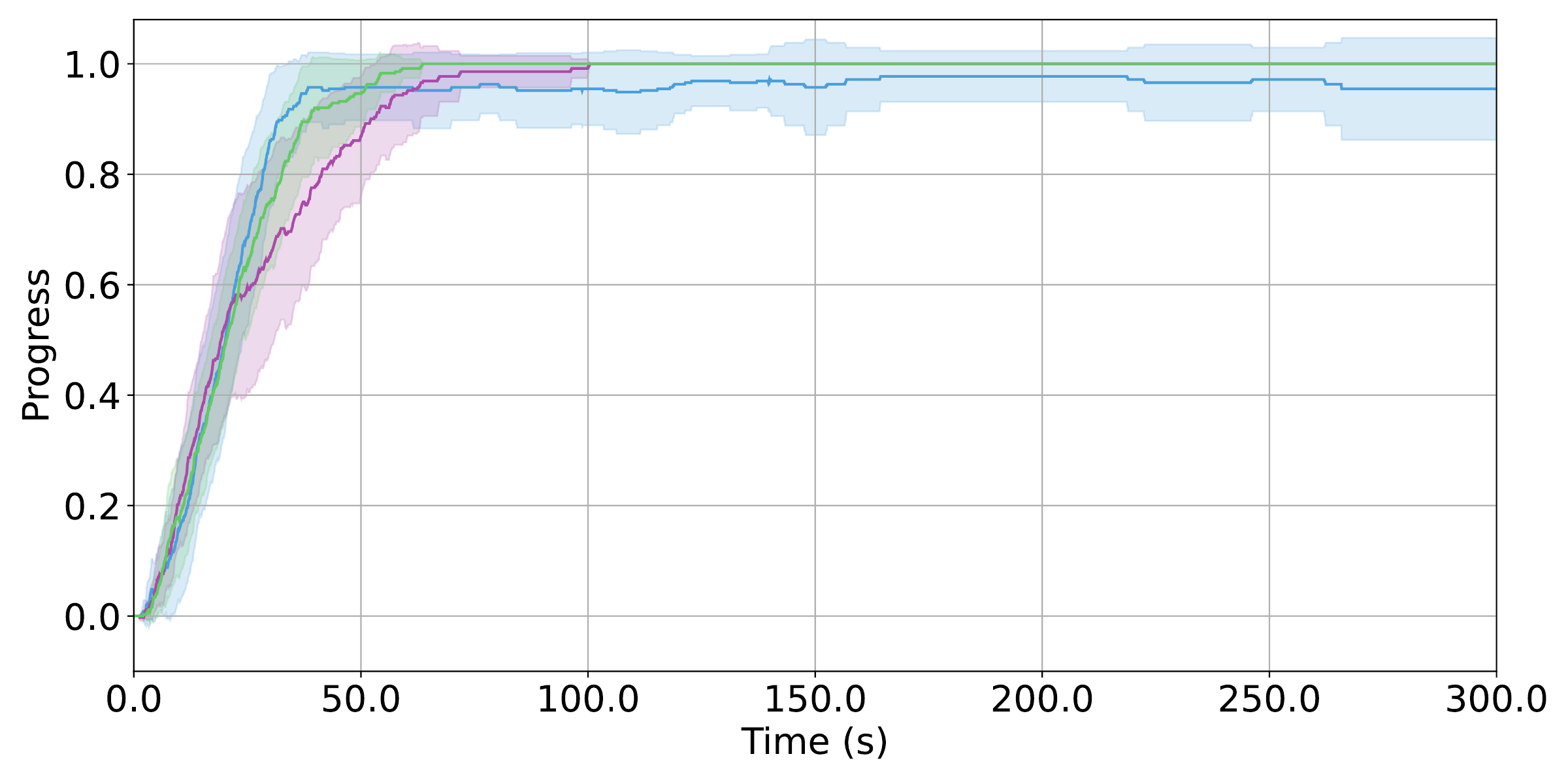}
    \caption{Puzzle 1; $O = 16$}
    \label{fig:sub1}
  \end{subfigure}\hfill
  \begin{subfigure}[t]{0.45\textwidth}
    \centering
    \includegraphics[width=\linewidth]{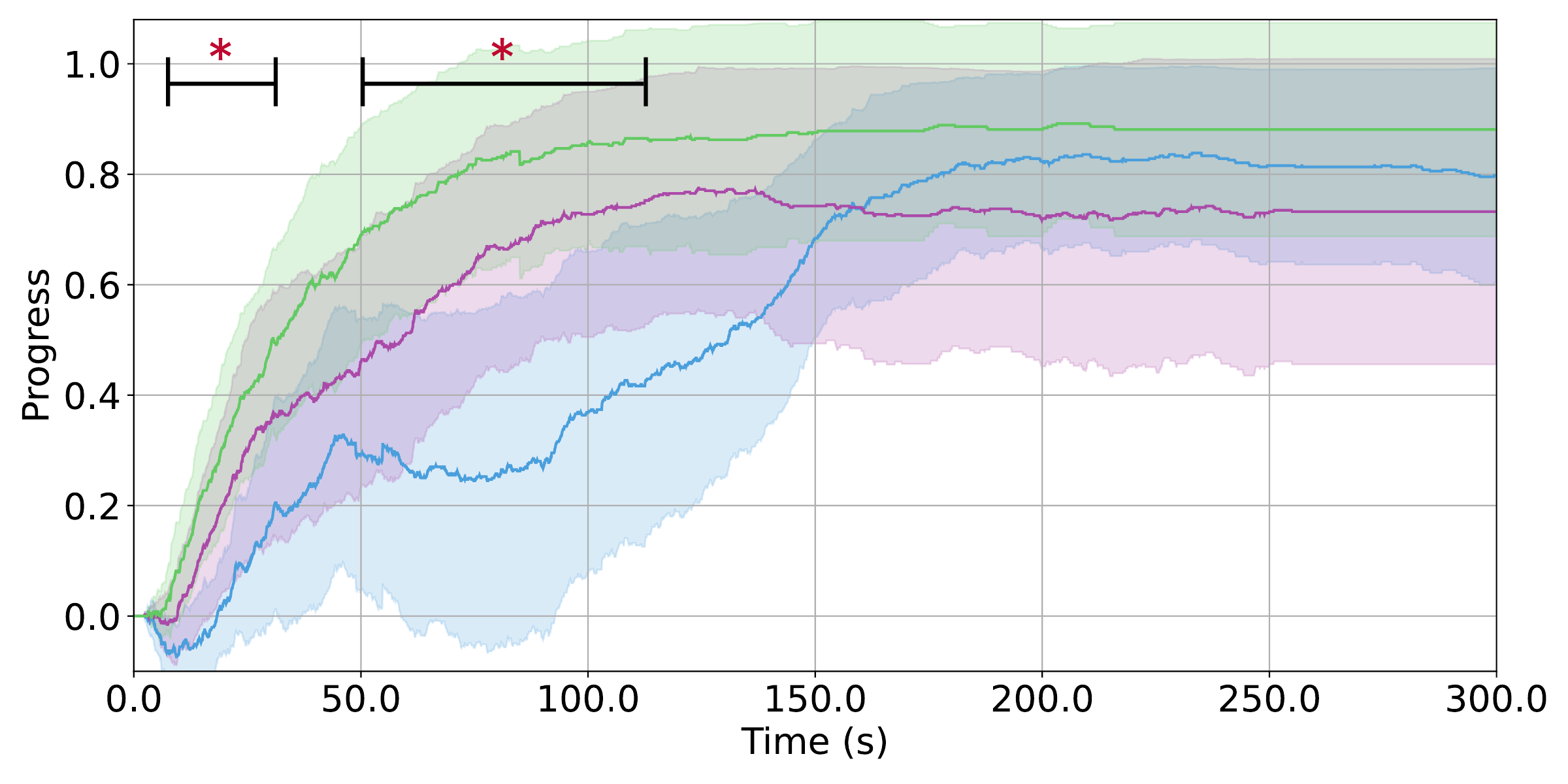}
    \caption{Puzzle 2; $O = 18$}
    \label{fig:sub2}
  \end{subfigure}\hfill

  \begin{subfigure}[t]{0.45\textwidth}
    \centering
    \includegraphics[width=\linewidth]{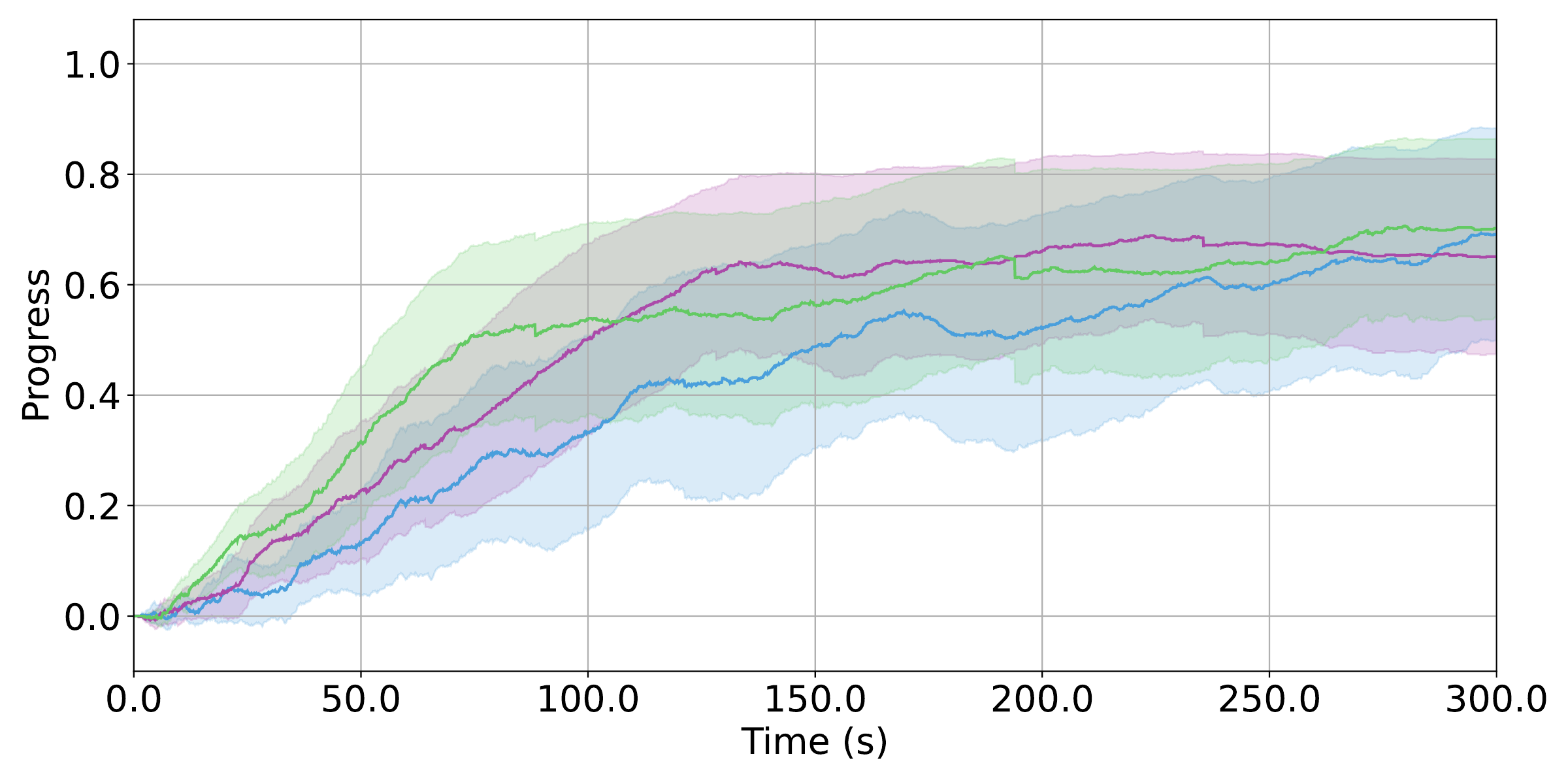}
    \caption{Puzzle 3; $O = 44$}
    \label{fig:sub3}
  \end{subfigure}\hfill
  \begin{subfigure}[t]{0.45\textwidth}
    \centering
    \includegraphics[width=\linewidth]{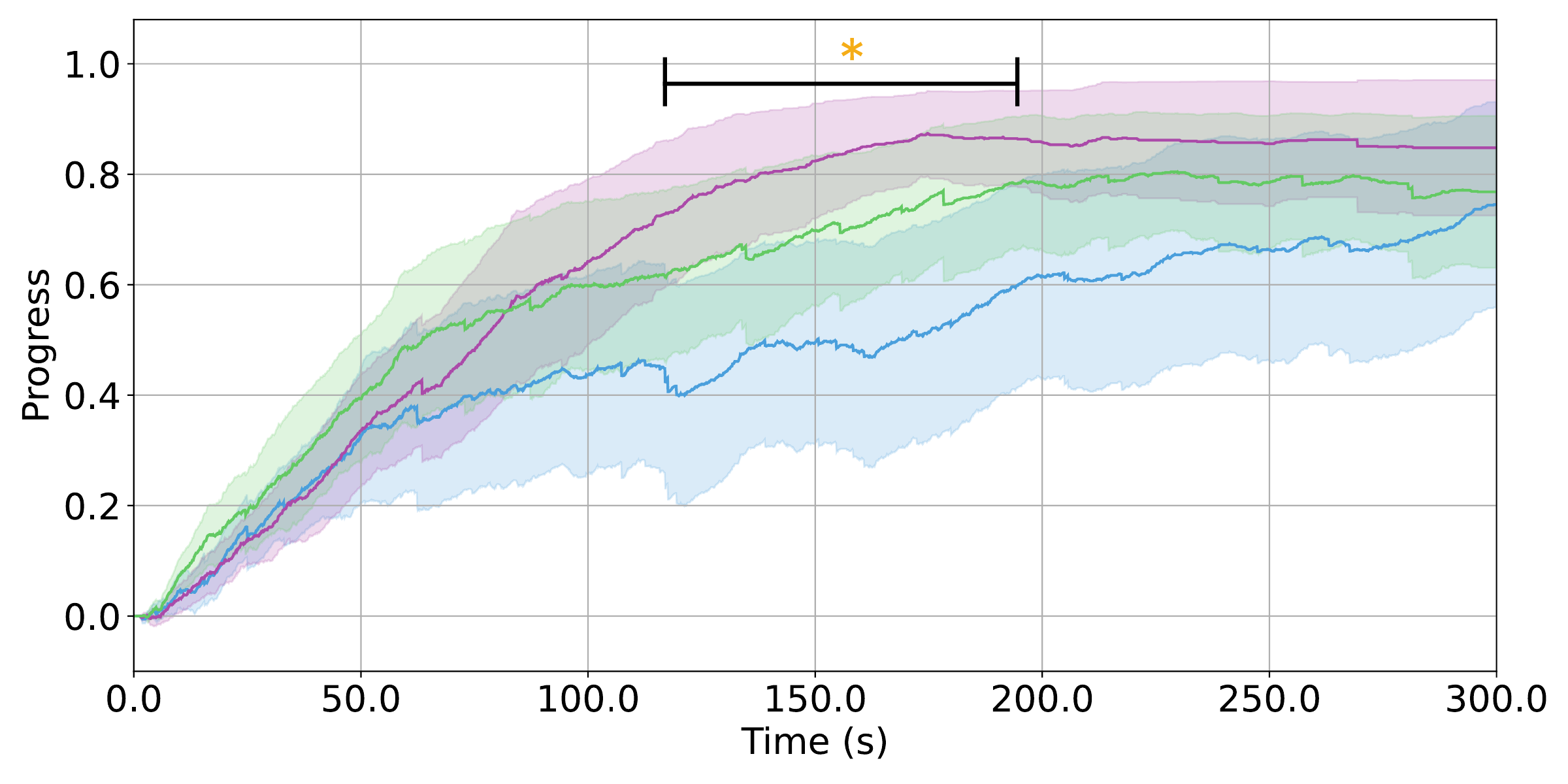}
    \caption{Puzzle 4; $O = 49$}
    \label{fig:sub4}
  \end{subfigure}

  \begin{subfigure}[t]{0.45\textwidth}
    \centering
    \includegraphics[width=\linewidth]{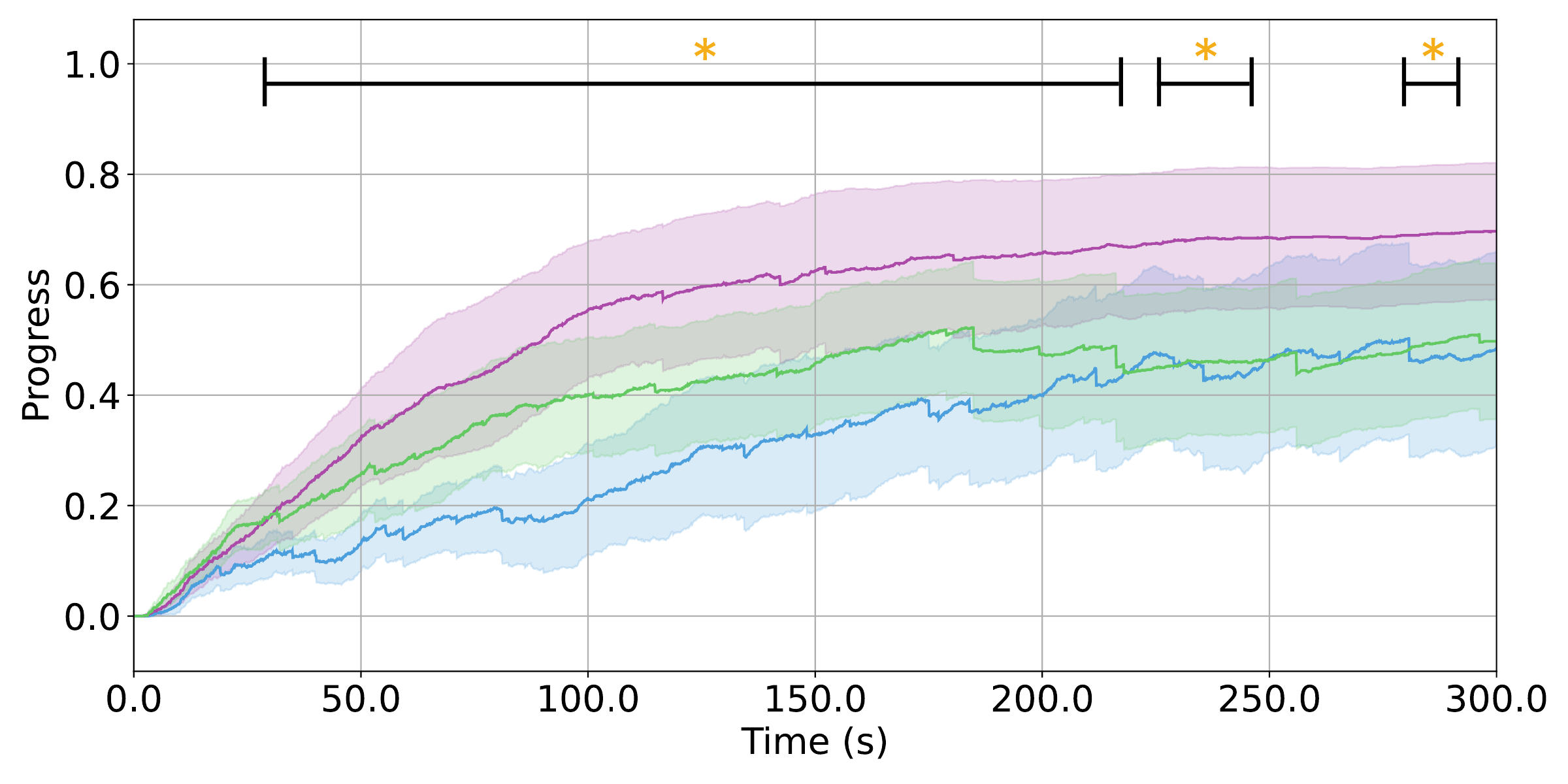}
    \caption{Puzzle 5; $O = 55$}
    \label{fig:sub5}
  \end{subfigure}\hfill
  \begin{subfigure}[t]{0.45\textwidth}
    \centering
    \includegraphics[width=\linewidth]{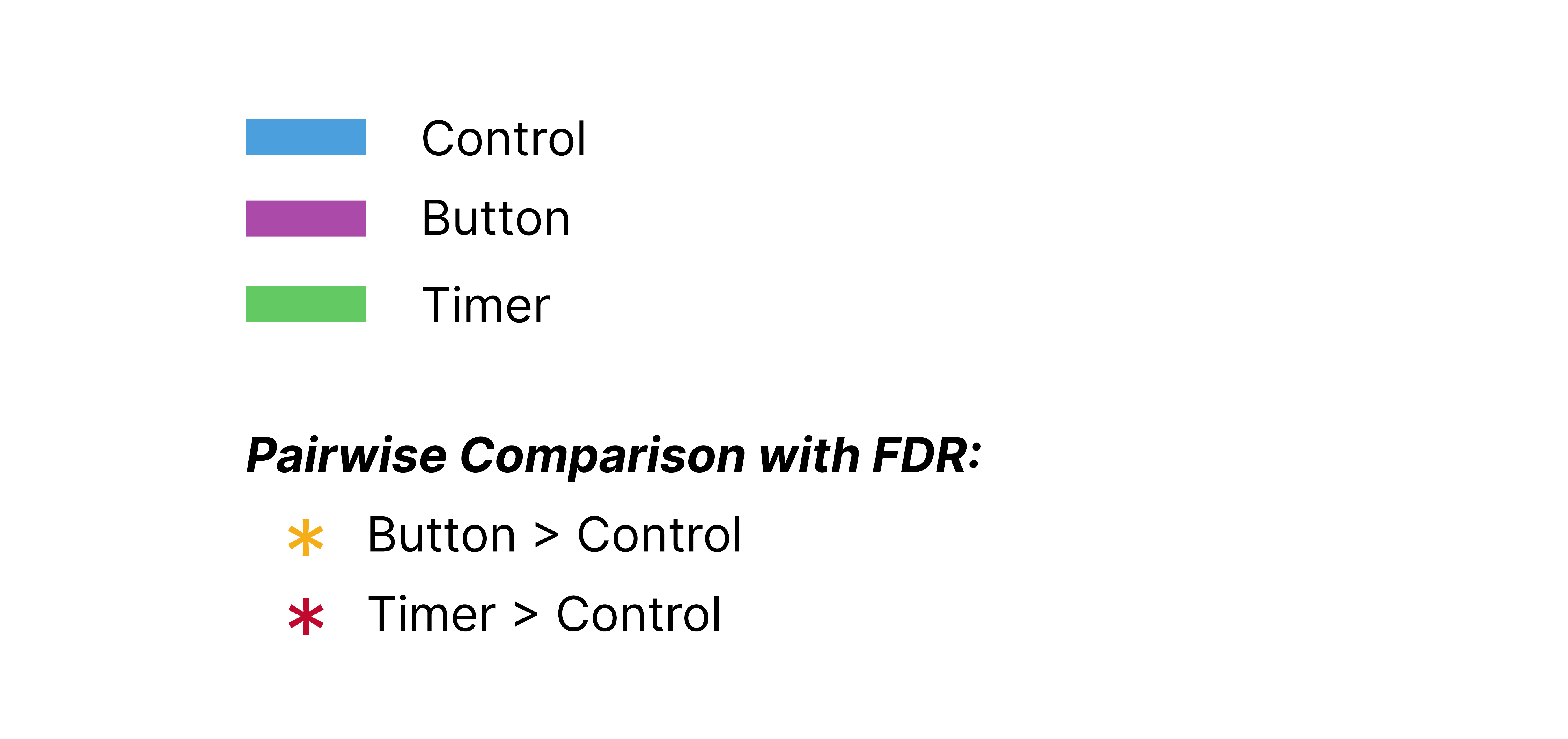}
    \label{fig:sub6}
  \end{subfigure}

  \caption{Average progress over time among all participants in each condition for the 5 puzzles in the main study phase. $O$ denotes the optimum number of moves to solve each puzzle from its initial state. Progress timelines were generated by first aligning individual progress over time to each participant's puzzle start time, and progress values were averaged across participants within each condition at every aligned time point. For participants who completed or exhausted their budget before 300 seconds, their final progress value was held constant to extend the trajectory, ensuring a full 300-second window for all participants. The shaded area indicates 95\% confidence intervals. Horizontal bars with asterisks indicate time intervals during which the difference between conditions was statistically significant based on t-tests with false discovery rate (FDR) correction.}
  \Description{The figure shows five line graphs comparing participants progress made over time on each puzzle three conditions: Control, Button, and Timer. In all puzzles, participants using AI assistance (especially the Button condition) progressed faster than those in the Control condition, particularly in the early to mid phases of the trial. Shaded regions indicate confidence intervals, and horizontal markers highlight time intervals where AI-assisted conditions significantly outperformed Control.}
  \label{fig:avg_progress_for_all_puzzles}
\end{figure*}

Although overall progress across conditions was similar, exploratory time-course analysis revealed nuanced differences for individual puzzles. In puzzle 2 (intermediate difficulty, $O = 18$), participants in the Timer condition showed significantly greater progress compared to those in the Control condition for most of the first 110 seconds. In puzzle 4 (advanced difficulty, $O = 49$), the Button condition outperformed the Control condition between 120 and 190 seconds. Similarly, in puzzle 5 (advanced difficulty, $O = 55$), the Button condition maintained a lead compared to the Control condition for most of the time from approximately 40 to 250 seconds, while also sparingly leading near the end until the budget was exhausted. See Figure~\ref{fig:avg_progress_for_all_puzzles} for progress over time for each puzzle.

\subsection{Behavioral Patterns}

While the two AI assistance modes differed in how help was triggered, we analyzed participants' behavioral patterns to understand how they engaged with each mode and how the differences may have shaped their interaction strategies.

\begin{figure*}
    \centering
    \includegraphics[width=0.99\linewidth]{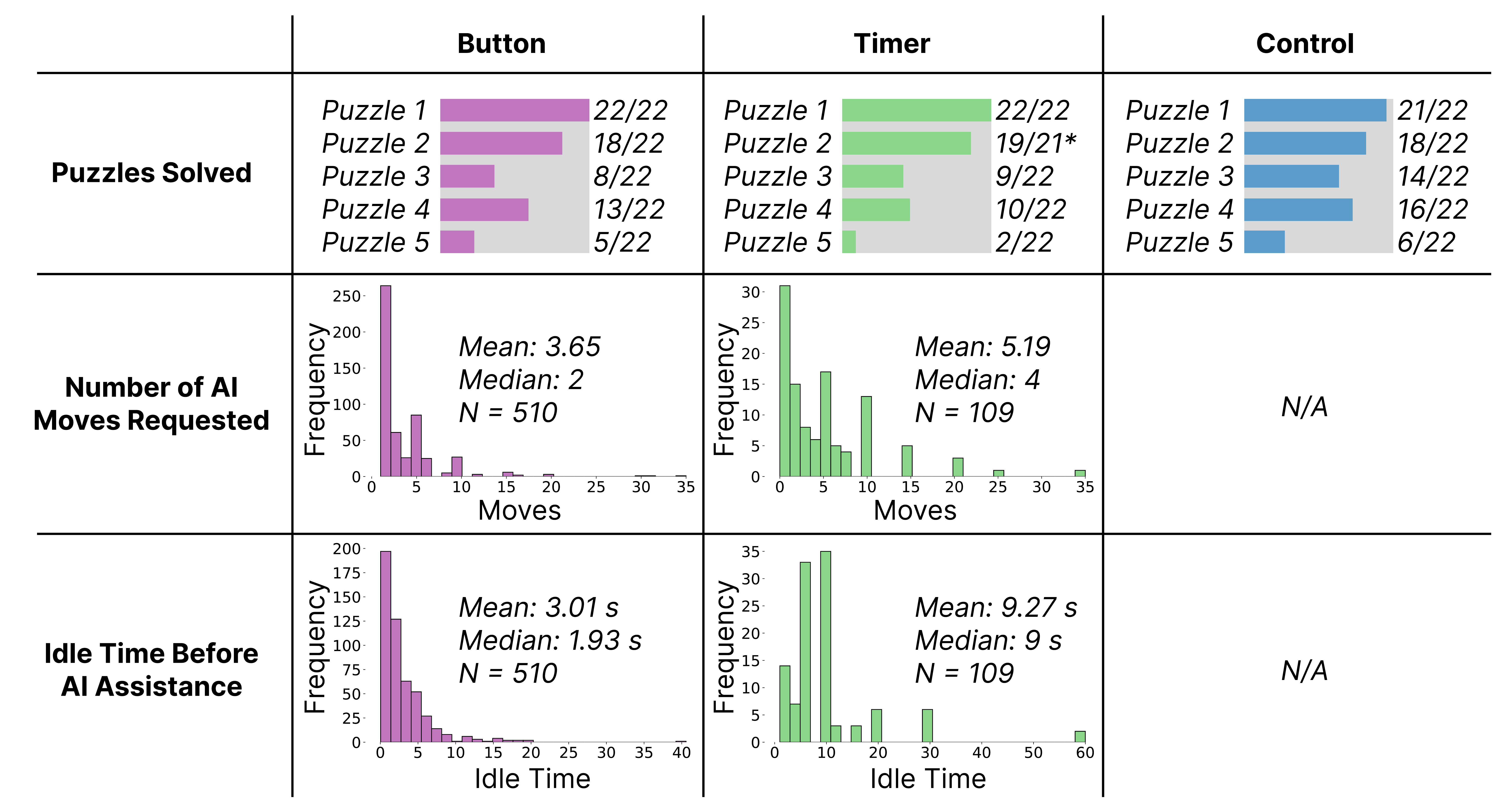}
    \caption{\textbf{Puzzles solved:} number of participants who solved the main study task puzzles across the three conditions. \textbf{Number of AI moves requested:} In the Button condition, this reflects every instance where AI assistance was requested ($N = 510$). In the Timer condition, it corresponds to the number of AI moves preset by users before starting each puzzle ($N = 109$). \textbf{Idle time before AI assistance:} In the Button condition, this is measured idle time before each AI assistance requested ($N = 510$). In the Timer condition, it reflects the user-defined idle time settings before starting each puzzle. \textit{Note: the mean values reported here differ from those in Section~\ref{idle_time}, which were calculated as per-puzzle averages only for puzzles where at least one AI assistance was triggered.} *One participant-puzzle pair was discarded due to missing data.}
    \Description{This figure compares puzzle completion, AI usage, and idle behavior across the Button, Timer, and Control conditions. Histograms for number of AI moves requested are presented for both Button and Timer conditions. Idle time before AI assistance is also presented in histograms for both the Button and Timer conditions.}
    \label{fig:AI_usage}
\end{figure*}

\subsubsection{AI Usage}

We analyzed the average number of AI-assisted moves in puzzles with at least one instance of AI assistance using a Gamma distribution with log link, including condition as a fixed effect and participant ID as a random intercept. Puzzles where no AI assistance was triggered, either because the user did not request help via the button in the Button condition or because the user's idle time never exceeded the preset idle threshold in the Timer condition, were excluded in this analysis, resulting in 77 valid puzzles in the Button condition and 75 valid puzzles in the Timer condition. No significant difference was found between the Button ($M = 24.182$, $SD = 14.186$) and the Timer ($M = 21.053$, $SD = 13.200$) conditions ($p=0.297$). Similarly, no difference was observed in the average number of AI help moves per request between the Button ($M = 6.454$, $SD = 6.490$) and the Timer ($M = 6.173$, $SD = 6.077$) conditions ($p = 0.556$). 

\begin{figure*}
    \centering
    \includegraphics[width=1\linewidth]{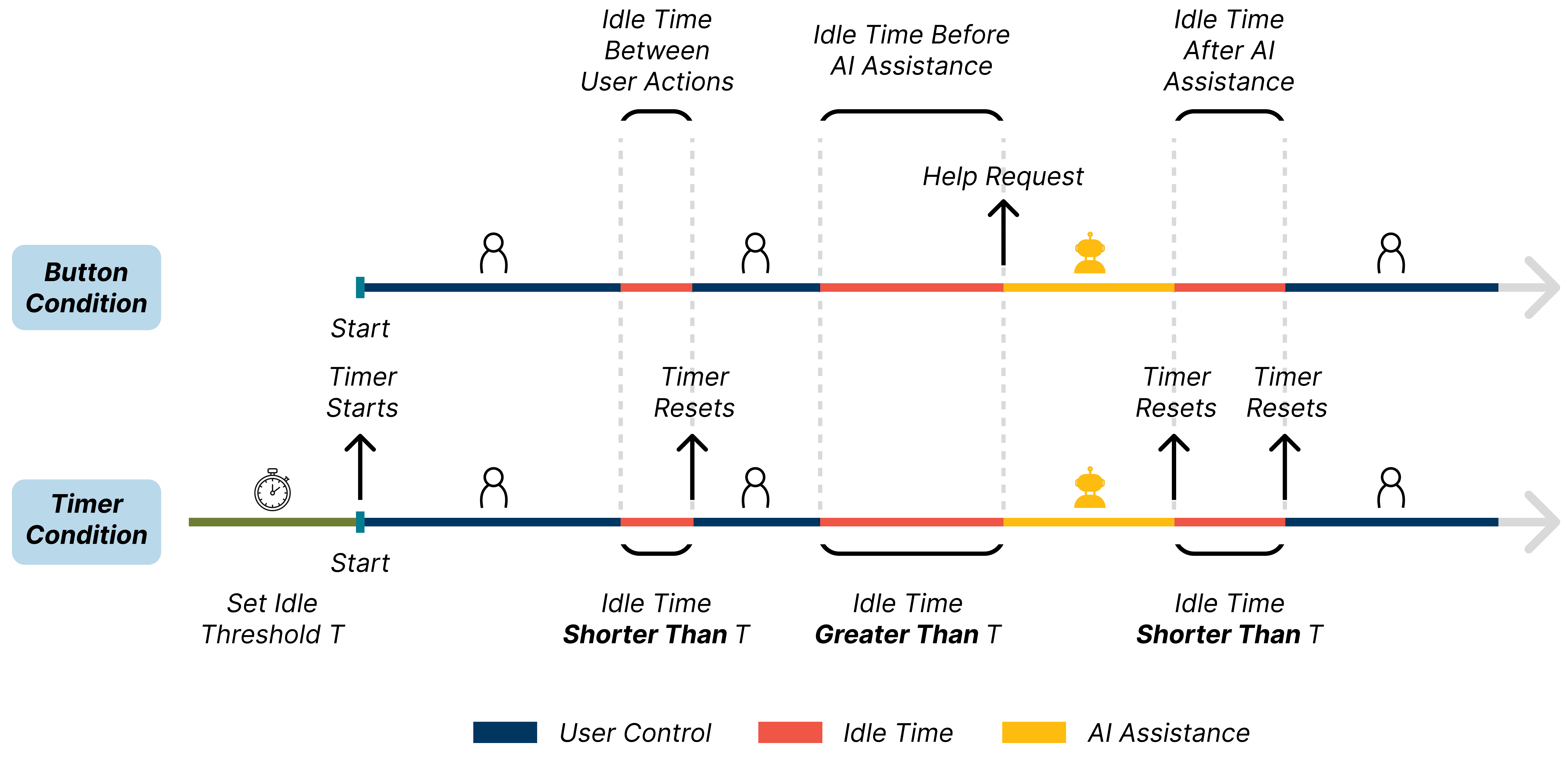}
    \caption{This figure illustrates simulated user flows in the two AI assistance conditions. \textbf{Button condition}: The user begins by moving puzzle pieces on their own (user control), followed by a brief idle period to think about next moves. After several more steps, the user feels stuck and idles for a longer duration. The user then decides to get some help from the AI assistance, so the user specifies the number of moves to help and clicks on the help Button, which triggers AI assistance. After the AI completes its moves, the user idles briefly to plan next moves before resuming puzzle-solving. \textbf{Timer condition}:
    Before starting the puzzle, the user configures two parameters: the idle time threshold (e.g., 10 seconds) and the number of moves the AI should provide when triggered (e.g., 5 moves). During gameplay, user action such as selecting or moving a puzzle piece resets the Timer, and no assistance is provided. However, if the user remains idle for a duration that exceeds the threshold, AI assistance is automatically triggered. The AI then executes the specified number of moves, after which the Timer resets. }
    \Description{This figure illustrates the simulated timelines of user behavior under the two AI assistance modes: Button and Timer. In the Button condition, the user starts by manually moving puzzle pieces until reaching a brief pause, followed by a longer idle period indicating they may be stuck. At that point, the user actively decides to request AI help by pressing a Help button and specifying the desired number of moves. The AI then takes over briefly, performing those moves. Once assistance ends, control returns to the user, who continues solving until future idle moments prompt additional help requests. In the Timer condition, the user predefines an idle-time threshold and number of moves for AI intervention before starting the puzzle. During gameplay, any user action resets the timer. If the user pauses but resumes within the threshold, no AI help is provided. However, if the user remains idle longer than the threshold, the AI automatically activates and performs the preset number of moves, after which the timer is reset.}
    \label{fig:user-flow}
\end{figure*}

\subsubsection{Idle Time Before and After AI Assistance}
\label{idle_time}
Idle time, the period during which participants took no action (selecting or moving a car), is a key behavioral measure in our study. Because the budget continuously decreased over time, participants were incentivized to remain engaged throughout the task. As a result, we interpret idle time as a period of deliberation, during which participants were thinking or planning their next move, rather than disengaging from the task. For this analysis, we included only puzzles with at least one instance of AI assistance (Button: $N = 77$; Timer: $N = 75$).

To understand how participants used the Button mode, we analyzed idle behavior preceding help requests. In this analysis, we considered idle time as the interval between any last changes to the selected car or the puzzle state (including those by the AI) and the moment AI assistance was activated. In the Button condition, for each participant–puzzle pair in the main study task, we measured the idle time before an AI request and found it to be longer than the idle times between other pairs of consecutive participant actions of sequential puzzle moves. This pattern held consistently across puzzles. The median idle time before an AI request exceeded on average $91.67\%$ ($SD = 11.49\%$) of the pauses between all user actions in the same participant-puzzle pair. See Figure \ref{fig:user-flow} for a visual illustration of idle time prior to an AI help request.

We then compared the idle times that preceded help requests between the Button condition and the Timer condition. We compared this by taking the average idle time before AI assistance for all participant-puzzle pairs that had AI assistance triggered at least once during that puzzle. We used a Gamma distribution with log link, using condition as a fixed effect and participant ID as a random intercept. Results show that the average idle time before AI assistance in the Timer condition ($M = 7.273s$, $SD = 5.437s$) is significantly longer than in the Button condition ($M = 4.661s$, $SD = 3.632s$) ($p = 0.009$).

We next examined how participants responded after receiving AI assistance. We measured the idle time following an AI assistance, defined as the duration between the end of an AI action to the closest next puzzle state change, which can be by either participant action or AI assistance requested or triggered. We used the same fixed effect and random intercept as that for idle before AI assistance. No difference was found between the idle time after AI assistance between the Button ($M = 4.394s$, $SD = 3.663s$) and the Timer ($M = 5.272s$, $SD = 3.521s$) conditions ($p = 0.164$).

\subsection{Perceived Task Load}
NASA-TLX responses were compared using the Mann-Whitney U test due to the violations of normality. Results show a significant difference in temporal demand ($p = 0.048$). Post-hoc pairwise comparisons suggest that participants experienced lower temporal demand in the Control condition ($M = 4.545$, $SD = 0.300$) compared to the Button condition ($M = 5.409$, $SD = 0.269$) and the Timer condition ($M = 5.545$, $SD = 0.244$). However, these differences did not reach statistical significance (Control vs. Button: $p = 0.111$; Control vs. Timer: $p = 0.069$). No significant differences were observed on other subscales of the NASA-TLX.

\begin{table*}[ht]
\centering
\caption{ANCOVA Results: Each category represents user perceptions of the AI from the post-study questionnaire, with an additional performance metric from the main study phase included as a covariate.}
\label{tab:ancova}
\resizebox{\textwidth}{!}{
\begin{tabular}{cl|cccc|cccc|cccc}
\toprule
\multirow{3}{*}{\textbf{Category}} & \multirow{3}{*}{\textbf{Mean (SD)}} & \multicolumn{4}{c}{\textbf{Total Help Used}} & \multicolumn{4}{c}{\textbf{Remaining Budget}} & \multicolumn{4}{c}{\textbf{Maximum Progress}} \\
\cmidrule(lr){3-6} \cmidrule(lr){7-10} \cmidrule(lr){11-14}
& & \multicolumn{2}{c}{\textbf{Condition}} & \multicolumn{2}{c}{\textbf{Covariate}} & \multicolumn{2}{c}{\textbf{Condition}} & \multicolumn{2}{c}{\textbf{Covariate}} & \multicolumn{2}{c}{\textbf{Condition}} & \multicolumn{2}{c}{\textbf{Covariate}} \\
\cmidrule(lr){3-4} \cmidrule(lr){5-6} \cmidrule(lr){7-8} \cmidrule(lr){9-10} \cmidrule(lr){11-12} \cmidrule(lr){13-14}
& & $p$ & $\eta^2$ & $p$ & $\eta^2$ & $p$ & $\eta^2$ & $p$ & $\eta^2$ & $p$ & $\eta^2$ & $p$ & $\eta^2$ \\
\midrule
\parbox[c][1.8cm][c]{2.0cm}{\centering \textbf{AI's Competence and Reliability}} & \parbox[c][1.8cm][c]{1.5cm}{\centering \textbf{Button:\\4.6 (1.7)\\Timer:\\5.3 (1.8)}} & 0.119 & 0.058 & \textbf{0.018*} & 0.128 & 0.225 & 0.036 & 0.767 & 0.002 & 0.117 & 0.059 & \textbf{0.013*} & 0.141 \\
\midrule
\parbox[c][1.8cm][c]{2.0cm}{\centering \textbf{AI's Helpfulness and Effectiveness}} & \parbox[c][1.8cm][c]{1.5cm}{\centering \textbf{Button:\\4.1 (1.9)\\Timer:\\5.0 (1.6)}} & \textbf{0.028*} & 0.112 & \textbf{0.005*} & 0.173 & 0.083 & 0.071 & 0.559 & 0.008 & \textbf{0.039*} & 0.100 & \textbf{0.024*} & 0.118 \\
\midrule
\parbox[c][1.8cm][c]{2.0cm}{\centering \textbf{Quality and Sufficiency of AI Offered Help}} & \parbox[c][1.8cm][c]{1.5cm}{\centering \textbf{Button:\\3.8 (1.9)\\Timer:\\4.8 (1.5)}} & \textbf{0.040*} & 0.099 & 0.102 & 0.064 & 0.066 & 0.080 & 0.996 & < 0.001 & \textbf{0.034*} & 0.105 & \textbf{0.043*} & 0.097 \\
\midrule
\parbox[c][1.8cm][c]{2.0cm}{\centering \textbf{AI's Interpretability}} & \parbox[c][1.8cm][c]{1.5cm}{\centering \textbf{Button:\\4.0 (1.8)\\Timer:\\4.5 (1.8)}} & 0.228 & 0.035 & \textbf{0.025*} & 0.117 & 0.371 & 0.020 & 0.704 & 0.004 & 0.238 & 0.034 & \textbf{0.029*} & 0.110 \\
\midrule
\parbox[c][1.8cm][c]{2.0cm}{\centering \textbf{Teamwork with AI}} & \parbox[c][1.8cm][c]{1.5cm}{\centering \textbf{Button:\\4.1 (1.9)\\ Timer:\\4.8 (1.6)}} & 0.075 & 0.075 & \textbf{0.001*} & 0.220 & 0.210 & 0.038 & 0.251 & 0.032 & 0.129 & 0.055 & 0.054 & 0.088 \\
\bottomrule
\end{tabular}
}
\end{table*}

\subsection{Post-Study Questionnaire}

To evaluate perceptions toward the AI between the two AI assistance modes, we first grouped questions into five categories: AI's competence and reliability, AI's helpfulness and effectiveness, quality and sufficiency of AI offered help, AI's interpretability, and teamwork with AI. See Appendix~\ref{post-study-questionnaire} for the full list of questions. We assessed internal consistency of each sub-category using Cronbach's alpha~\cite{cronbach1951coefficient}: AI's competency and reliability ($\alpha = 0.966$), AI's helpfulness and effectiveness ($\alpha = 0.906$), quality and sufficiency of AI offered help ($\alpha = 0.873$), AI's interpretability ($\alpha = 0.864$), and teamwork with AI ($\alpha = 0.933$). Using the commonly recommended reliability threshold of $0.8$ \cite{nunnally1994psychometric}, these values confirm strong internal consistency across all sub-categories.

Results reveal marginal differences between the two conditions in the AI's helpfulness and effectiveness (Mann-Whitney test: $p = 0.097$, Rank-biserial correlation $= 0.293$) and the quality and sufficiency of AI offered help (independent t-test: $p = 0.063$, Cohen's d $ = -0.576$). In both cases, the Timer condition received higher ratings. To further probe these differences, we conducted an analysis using ANCOVA with condition as a fixed effect and controlling for task performance metrics during the main study phase as covariates, including only one covariate per model. This approach allowed us to examine how perceptions of the AI differed between assistance modes while controlling for these variables.

\subsubsection{AI's Competence and Reliability}
Participants in the Timer condition reported higher ratings ($M = 5.273$, $SD = 1.804$) than those in the Button condition ($M = 4.606$, $SD = 1.711$), though this difference was not statistically significant ($p = 0.210$). ANCOVA analysis revealed two significant covariates: the total number of times help was used ($p = 0.018$) and maximum progress in the main study phase ($p = 0.013$), both of which had a significantly positive influence on ratings.

\subsubsection{AI's Helpfulness and Effectiveness}
A trend toward higher ratings was observed in the Timer condition ($M = 5.045$, $SD = 1.616$) compared to the Button condition ($M = 4.080$, $SD = 1.909$), although this difference did not reach statistical significance ($p = 0.097$). ANCOVA analysis revealed that when controlling for either the total number of times help was used or maximum progress in the main study phase, ratings in the Timer condition were significantly higher than those in the Button condition ($p = 0.028$ and $p = 0.039$ respectively). Both covariates also showed significant positive effects on ratings ($p = 0.005$ and $p = 0.024$ respectively).

\subsubsection{Quality and Sufficiency of AI Offered Help}
A marginal difference was observed between the Button ($M = 3.795$, $SD = 1.867$) and the Timer ($M = 4.784$, $SD = 1.549$) conditions ($p = 0.063$). ANCOVA analysis revealed that when controlling for the total number of times help was used, ratings in the Timer condition were significantly higher than those in the Button condition ($p = 0.040$). Similarly, when controlling for maximum progress in the main study phase, the Timer condition again received significantly higher ratings than the Button condition ($p = 0.034$). Maximum progress itself also emerged as a significant covariate ($p = 0.043$).

\subsubsection{AI's Interpretability}
No significant difference was observed in interpretability ratings between the Button condition ($M = 3.985$, $SD = 1.798$) and the Timer condition ($M = 4.485$, $SD = 1.778$). However, both the total number of times help was used and the maximum progress achieved during the main study phase emerged as significant covariates ($p = 0.025$ and $p = 0.029$ respectively), each positively associated with higher ratings.

\subsubsection{Teamwork with AI}
No significant difference was found in the team work ratings between the Button condition ($M = 4.055$, $SD = 1.924$) and the Timer condition ($M = 4.755$, $SD = 1.608$). However, the total number of times help was used in the main study phase emerged as a significant covariate ($p = 0.001$) with a positive correlation with the ratings.

\subsubsection{Influence of Prior AI Experience and ATI}
We examined whether prior AI experience or ATI scores influenced participants' subjective ratings of the AI. Prior AI experience, measured as a four-level categorical variable, was analyzed using ANOVA and showed no significant effects on any of the subjective rating measures. ATI scores were analyzed using Pearson's correlations and likewise showed no significant associations with any measures (all $p > 0.05$).

\section{Qualitative Results}

The first three authors analyzed responses to the open-ended questions using framework analysis \cite{gale2013method}. Analysis began with familiarization with the data through active reading and re-reading of user responses. An initial coding framework was then developed using a hybrid deductive-inductive approach; the coded data were subsequently interpreted by identifying patterns, relationships, and overarching themes. Coding was conducted at the semantic level, and the framework was iteratively refined throughout the analysis process. The following section presents our findings.

\subsection{Competence and Helpfulness of the AI}

Participants from both conditions generally perceived the AI as competent and reliable, with many noting that it provided logical and effective moves. For instance, P32 (Button) remarked that \textit{``it knew where the next move was,''} while P54 (Timer) commented that \textit{``it was swift and reliably progressed the puzzle when I couldn't.''} Participants valued the efficiency of the AI's assistance. P45 (Timer) described the AI as \textit{``efficient''} and P60 (Timer) noted that \textit{``it was efficient with the moves that it had to make.''}

However, in both the Button ($N = 5$) and Timer ($N = 3$) conditions, a few participants expressed reservations regarding the AI's utility in puzzle solving, noting that some of the AI's moves appeared either meaningless or random. For instance, P33 (Button) felt that the AI \textit{``did not really help as it just moved the blocks without providing the solution.''} Similarly, P24 (Button) commented, \textit{``in my opinion, the moves just seemed to be random without much meaning behind them.''} Another participant, P27 (Button), noted that the AI's interventions were often redundant: \textit{``It typically just redid moves I'd already made when it intervened, so not very useful.''} Similar concerns were expressed in the Timer condition as well. P64 (Timer) reported, \textit{``to me it looked like it was wasting moves because they seemed pointless.''} P50 (Timer) concluded that AI \textit{``was not helpful at all.''}

\subsection{Control and Timeliness of Assistance}
Across the two AI assistance modes, participants reported similar levels of controllability, with subtle differences in how they perceived the AI’s involvement. In the Button condition, participants emphasized the value of control. P41 (Button) remarked, \textit{``I liked how I was able to control how many moves it made,''} and P23 (Button) similarly noted, \textit{``It gave the option to choose how many steps I wanted AI to assist with.''} 
Participants in the Timer condition also valued being able to regulate when and how the help was delivered. P64 (Timer) explained that they valued \textit{``that you could choose at what intervals they [AI] would help you,''} and P66 (Timer) similarly liked \textit{``being able to customize when it would jump in, along with how many steps it would take.''} Beyond this sense of control, some participants in the Timer condition perceived the assistance as proactive and timely. P61 (Timer) appreciated that \textit{``the AI stepped in when I was stuck and offered logical moves that helped me progress. It felt like a gentle nudge rather than taking over completely.''} Likewise, P58 (Timer) noted that they liked \textit{``that it came in and assisted when I was stuck.''}

\subsection{Need for Explanation}
Multiple participants ($N = 5$) emphasized the need for explanations as a key improvement. When asked: ``What did you dislike about the AI?'', the difficulty in discerning the rationale behind AI moves was a common issue. P44 (Button) responded, \textit{``I couldn't understand its logic.''} Similarly, P51 (Timer) noted, \textit{``sometimes you didn't understand why they did something,''} and P42 (Button) expressed that they \textit{``couldn't tell where it was going with some moves.''} The absence of explanations often confused participants and occasionally made them suspicious that the AI was intentionally making poor choices. For instance, P41 (Button) commented, \textit{``Sometimes I felt like it was taking too long, and would choose incorrect moves on purpose.''} Similarly, P56 (Timer) stated that, \textit{``I didn't like that I couldn't always tell if it was helping me or not.''} 

When asked what the AI should do differently, participants repeatedly requested more interpretability. P45 (Timer) suggested \textit{``commentary with its moves - i.e., explaining why it did what it did so you can pick up where it left off.''} Likewise, P34 (Button) proposed that \textit{``the AI could provide clear explanations for its suggested moves.''} as this \textit{``would help build trust and understanding.''}

\subsection{Desired AI Behavior}
Participants offered a range of suggestions for modifying the AI's behavior to their liking. A recurring theme across both conditions was the speed, with many suggesting that the AI should move faster. P26 (Button) and P28 (Button) both criticized the AI as \textit{``not fast enough''} and \textit{``too slow.''} Similarly, P65 (Timer) commented, \textit{``I expected the AI to be fast in its movements.''}

Beyond speed, participants expressed preferences for features of the alternate, but unknown to them, condition. In the Button condition, participants suggested that the AI should act more proactively based on the context. P23 (Button) suggested that the AI should \textit{``offer a suggestion after you haven't made a decision in a certain time period.''} P41 (Button) expected that the AI should \textit{``know when I would need it to activate [it], and have a mind of itself. So if I was struggling the AI could step in.''} Conversely, one participant (P59, Timer) desired greater manual control, noting that it was \textit{``hard to draw a line between asking for some help and it doing it for you - would have been better to have been deployed by a Button.''}

\subsection{No-Assistance Problem-Solving Behavior}

Participants from the Control condition ($N = 22$) reported two main approaches to solving the puzzles: trial-and-error ($N = 9$) and systematic planning ($N = 12$). The trial-and-error approach involved making rapid moves with minimal planning, testing different configurations to see what works. The systematic planning approach involved backtracking the red car’s position, moving larger blocking cars first, and planning several moves ahead, sometimes working backward from the goal state. One participant (P17) reported no systematic strategy and expressed frustration at failing to solve the puzzles.

\paragraph{Preference for assistance from participants in the Control condition}

When asked about preferred type of help, most participants ($N = 15$) requested small nudges or contextual clues at critical junctures, often limited to a single move. Specific requests included the option to see the first move (P3) or receiving hints when \textit{``end of longer pieces would block the path.''} (P21). P1 desired a limited number of hints, stating that hints \textit{``take away the satisfaction of solving [the puzzle].''} Conversely, one participant (P12) expressed a desire to delegate the entire puzzle to the AI when stuck, and another requested the puzzle constraints to be relaxed to mitigate difficulty and frustration. A participant (P19) who used a trial-and-error strategy indicated that high-level strategic guidance would have been helpful. 

\paragraph{Desired AI role}

A majority of participants ($N = 17$) expressed a preference for the AI to function as a tool. P16 justified this choice by emphasizing a desire to avoid the \textit{``distraction of being a teammate,''} which they associated with excessive communication, and a preference for a basic utility that provides moves. Another participant (P11) preferred the AI as a tool because they did not want the AI to \textit{``act like a human''} by \textit{``yelling at [them] about [their] performance.''}

In contrast, few participants expressed that it would help them learn and improve if AI functioned like a coach. P6 stated that \textit{``AI should be like a coach who wanted to make player a better planner next time to play successfully,''} and P19 preferred AI to be \textit{``like a coach because I can grow.''}

Irrespective of the preferred AI role, participants expressed a strong desire for control over the puzzle-solving process, estimating they should maintain above 80\% control over the moves to \textit{``feel a sense of accomplishment''} (P3).

\paragraph{Assistance timing}

When asked about the timing of assistance, participants expressed differing preferences based on their solving strategy. The majority of participants who planned systematically ($N = 10/12$) indicated that they preferred on-demand help. In contrast, participants who relied on trial-and-error ($N = 5/9$) reported a preference for contextualized help triggered by failed attempts. One participant (P19) preferred no assistance, stating \textit{``I do not need AI to help as long as there is a solution.''}

\section{Discussion}

Our study compared two AI assistance delivery modes where assistance was triggered either explicitly through a button press (Button mode) or automatically after a user-defined inactivity period (Timer mode), along with a control condition. Task progress by participants was similar across the two AI assistance modes and the control group. Despite similar performance, participants in the Timer condition reported more positive impressions of the AI. In the following section, we discuss our findings and their potential implications.

\subsection{AI Assistance vs. No Assistance}

\paragraph{Resource allocation}

Move precision was significantly enhanced under AI-supported conditions compared to the no-assistance control condition. This effect is attributable to the algorithmically precise AI moves, which mitigated the inherent variability of unaided human performance. Although participants with AI support showed momentary leads in progress, AI assistance did not translate into statistically superior overall task performance, a pattern consistent with prior findings \cite{wu2025human}. While this may have been a result of the specific experimental cost structure implemented, it highlights a challenge in optimal real-time resource allocation between human effort (time) and costly AI assistance. Theoretically, maximizing progress required continuous evaluation of the trade-off between independent problem-solving and AI assistance. Independent solving was efficient only when its rate of progress exceeded the marginal benefit provided by costly AI assistance. Participants who were unable to optimize this resource allocation incurred higher costs without commensurate gains in overall progress relative to the control condition. Such suboptimal resource allocation may reflect broader challenges in human-AI collaboration where decision-makers fail to accurately calibrate when to request assistance versus independent effort for maximizing overall task outcomes~\cite{steyvers2024three, vaccaro2024combinations}.

\paragraph{User perception and automation bias}

Our subjective results reveal a positive correlation between AI usage frequency and favorable perceptions of the AI's competency and helpfulness. These positive impressions appear to have reinforced continued AI engagement, as participants prioritized perceived progress gains over the expenses incurred, thereby increasing overall costs. This behavioral pattern suggests a self-reinforcing cycle where initial success with AI assistance may have strengthened trust and satisfaction, which in turn promoted further reliance. While this behavior does not constitute classic automation bias, where users uncritically defer to automated decisions~\cite{parasuraman1997humans}, it reflects a trust-driven overreliance dynamic that aligns with mechanisms underlying automation bias.

\paragraph{AI as a tool vs partner}

Qualitative data indicates that many users framed AI not as a collaborative partner but as an intelligent, reliable instrument expected to engage judiciously and contextually. This finding suggests that the relational framing of AI is context-dependent. In high-efficiency, task-oriented scenarios, utility is prioritized over relational support, contrasting with findings in more relational and affective settings~\cite{pataranutaporn2025my, pataranutaporn2021ai}. However, even within the instrumental framing of AI, there is no uniform acceptance of AI assistance. Subjective feedback, in our study, ranged from perceiving the AI's actions as logical and effective to expressing confusion and distrust regarding its competency and intentions. While some of the distrust may be attributable to factors such as the Out-of-the-Loop performance problem~\cite{endsley1995out}, the lack of clear explanations of the AI's actions likely amplified user detachment, overriding the mitigating effect of short intervention intervals. This indicates that for AI to function as a ``reliable instrument,'' interpretability is required in addition to timely intervention.

\subsection{AI Assistance Delivery Mechanisms and User Perceptions}

Despite the differences in how AI assistance was triggered, participants in both conditions converged on similar task performance and overall AI usage. This convergence may be explained through bounded rationality and satisficing~\cite{simon1955behavioral}, which posit that individuals make decisions under cognitive and informational constraints by relying on simple, ``good-enough'' rules rather than exhaustive optimization. Under time pressure and explicit monetary costs, participants were unlikely to compute an optimal help-seeking strategy and instead relied on simple heuristics such as requesting assistance when progress stalled. Given comparable skill levels, participants likely encountered impasses at similar points in the task. In the Button condition, these impasses triggered explicit help requests, whereas in the Timer condition they manifested as longer pauses that activated inactivity-triggered assistance. As a result, participants in both conditions may have relied on the same underlying ``stuckness''-based decision rule, leading them to satisfice toward a cost-efficient equilibrium of AI usage and yielding similar performance across conditions.

Although the two AI assistance modes resulted in similar performance and budget efficiency, they produced distinct user experiences. After controlling for assistance frequency and maximum progress achieved, Timer condition participants rated the AI as significantly more helpful and supportive. This divergence suggests that delivery mode itself shaped participant satisfaction independent of the actual level of assistance received. Participants in the Timer condition characterized the AI's intervention as timely, noting that it ``stepped in'' when needed and felt like a supportive presence. These findings indicate that delivery mode modulates users' perceptions of the AI's collaborative intent: identical assistance can be perceived as appropriately supportive when well-timed, or as insufficient or disruptive when mistimed. Thus, assistance delivery mode constitutes a critical design factor in shaping user experience of mixed-initiative systems.

While the two AI assistance delivery modes were designed to achieve the same functional outcome, they elicited different cognitive and experiential trade-offs. From the perspective of adjustable autonomy~\cite{bradshaw2003dimensions}, which characterizes how initiative and control are flexibly shared between users and automated systems, the Button and Timer modes represent distinct allocations of initiative over assistance activation. In the Button mode, users had to actively decide when to request assistance, requiring additional cognitive resources for monitoring and self-evaluating their need for help. In contrast, the Timer mode may have allowed participants to achieve a balance between convenience and comfort through automated assistance activation, allowing initiative to be partially delegated to the system, but mediated by user-defined idle-time thresholds. This may have made the timing of system actions feel more appropriate and less effortful than requesting help moment-to-moment, which may also explain why participants in the Timer condition reported more positive attitudes toward the AI.
Despite reduced direct control over the immediate activation of AI assistance in the Timer condition, participants did not report a loss of agency. This suggests that perceived autonomy may depend on both explicit control mechanisms and a user's ability to position the system at a personally acceptable level of delegation. Delivering assistance at appropriate moments that feels timely, even if triggered automatically based on user settings, may preserve a sense of autonomy. This distinction between actual and perceived control offers an important nuance for designing mixed-initiative systems that support agency without demanding constant decision-making.

\subsection{Design Implications}

\paragraph{Supporting calibrated assistance decisions}

Our quantitative results on remaining budget suggest that participants had difficulty optimizing the cost-benefit trade-offs between independent problem-solving and costly AI assistance. This indicates that systems should also support users in the meta-decision of when to engage AI assistance in order to improve overall efficiency. Systems could potentially monitor user behavioral cues, such as gaze patterns, emotional state, and interaction behaviors, to identify optimal assistance timing based on observable signals linked to users' internal states and performance~\cite{bixler2015automatic, nargund2025understanding}.

\paragraph{Configurable proactivity}

The combination of the user's desire for control and their expectations of AI acting contextually suggests that future mixed-initiative systems should support configurable proactivity by enabling users to specify AI intervention triggers based on task-specific conditions and contextual cues. Importantly, such configuration mechanisms should be based on observable and human-interpretable parameters, such as inactivity periods or specific task states, to align automated assistance with the user's mental models of help-seeking.

\paragraph{Automation bias mitigation in costly assistance contexts}

When AI assistance is both precise and costly, positive perceptions of competence can drive increased use even when it fails to improve overall task efficiency. Our results indicate that favorable impressions of AI-supported progress can reinforce continued engagement, increasing costs without yielding statistically superior performance. Designers should therefore anticipate self-reinforcing overuse in efficiency-oriented tasks and provide feedback that makes the long-term cost–performance trade-offs visible.

\paragraph{Make AI actions interpretable to support seamless handoffs}

In addition to the timing of assistance, a user's understanding of AI rationale is critical for calibrating trust and facilitating handoffs. Our qualitative findings indicate that while some participants perceived the AI's moves as strategically sound, others found them arbitrary, potentially making it harder for them to resume solving after the assistance ended. This suggests that exposing individual AI actions alone may be insufficient for interpretability in sequential problem-solving tasks. Designers should consider additional cues that communicate the AI's higher-level intent, such as the intermediate goal of a sequence of moves or long-term strategy, to support sensemaking and enable smoother handoffs.

\section{Limitations \& Future Work}

While our findings indicate that assistance delivery mode impacts user experience in collaborative problem-solving, several limitations should be considered when interpreting and generalizing these results.
We used Rush Hour puzzles as our testbed because they capture several important features of real-world collaborative problem-solving, including time pressure, sequential decision-making, strategic planning, and opportunities for help-seeking. However, puzzles are fundamentally well-defined problems with explicit goals and structured solutions. Consequently, our findings may not directly translate to more open-ended or creative tasks where problem definitions are ambiguous, multiple valid approaches exist, or success criteria are subjective. Future work should systematically examine whether inactivity-triggered assistance maintains its experiential benefits across a broader range of task types. 
Our study captured user perceptions and preferences during a single session with limited exposure to each assistance mode. While this design allows us to isolate immediate experiential differences, it does not account for how preferences and behaviors might evolve with extended use. Users may habituate to initially preferred modes, discover new strategies that change their assistance needs, or recalibrate their expectations as they gain experience with the task or the AI's capabilities. Longitudinal studies are needed to assess whether these preferences persist over time and whether users develop more sophisticated strategies for leveraging different assistance modes. Understanding these temporal dynamics would inform the design of assistance systems that remain effective and satisfying throughout extended use.
We compared two specific assistance modes that represent important points in the design space of human-AI collaboration. However, many other timing mechanisms could be designed. For example, adaptive systems might dynamically adjust intervention timing based on learned user preferences, physiological signals, or performance patterns. Scheduled assistance that triggers at fixed intervals, context-aware systems that consider task phase or environmental factors, or hybrid approaches that combine multiple triggering mechanisms may offer different experiential qualities. While our findings suggest that timing modality matters for user experience, further research is needed to characterize the full design space and identify which mechanisms work best for different contexts and user populations.

\section{Conclusion}
Determining when AI should intervene in collaborative problem-solving remains a persistent challenge in mixed-initiative systems. While existing approaches emphasize optimizing the timing of assistance, our work reveals that the mechanism through which assistance is delivered, independent of its impact on task outcomes, can shape user experience.
For designers of collaborative AI systems, our finding suggests that inactivity-triggered assistance offers a practical pattern for balancing user control with reduced cognitive burden. Rather than treating assistance timing as a secondary implementation detail, designers should consider delivery modality as a first-class design concern that directly impacts user agency and engagement. As AI systems become increasingly capable, understanding and optimizing how they integrate into human workflows will be essential for creating human-centered collaborative experiences that users find both effective and satisfying.

\section{GenAI Usage Disclosure}
LLMs were used to explore potential data analysis approaches and to suggest recommendations for improving text phrasing and clarity. Any suggestions provided by LLMs served only as input to inform our own decision-making. AI tools were also employed during study interface development for code debugging.



\bibliographystyle{ACM-Reference-Format}
\bibliography{main}

@article{kobis2025delegation,
  title={Delegation to Artificial Intelligence can increase dishonest behaviour},
  author={K{\"o}bis, Nils and Rahwan, Zoe and Rilla, Raluca and Supriyatno, Bramantyo Ibrahim and Bersch, Clara and Ajaj, Tamer and Bonnefon, Jean-Fran{\c{c}}ois and Rahwan, Iyad},
  journal={Nature},
  pages={1--9},
  year={2025},
  publisher={Nature Publishing Group UK London},
  url={https://doi.org/10.1038/s41586-025-09505-x},
  doi={10.1038/s41586-025-09505-x}
}

@inproceedings{erlei2022s,
author = {Erlei, Alexander and Das, Richeek and Meub, Lukas and Anand, Avishek and Gadiraju, Ujwal},
title = {For What It’s Worth: Humans Overwrite Their Economic Self-interest to Avoid Bargaining With AI Systems},
year = {2022},
isbn = {9781450391573},
publisher = {Association for Computing Machinery},
address = {New York, NY, USA},
url = {https://doi.org/10.1145/3491102.3517734},
doi = {10.1145/3491102.3517734},
booktitle = {Proceedings of the 2022 CHI Conference on Human Factors in Computing Systems},
articleno = {113},
numpages = {18},
location = {New Orleans, LA, USA},
series = {CHI '22}
}

@book{lazar2017research,
  title={Research methods in human-computer interaction},
  author={Lazar, Jonathan and Feng, Jinjuan Heidi and Hochheiser, Harry},
  year={2017},
  publisher={Morgan Kaufmann},
  isbn={978-0-12-805390-4}
}

@ARTICLE{allen1999mixed,
  author={Allen, J.E. and Guinn, C.I. and Horvtz, E.},
  journal={IEEE Intelligent Systems and their Applications}, 
  title={Mixed-initiative interaction}, 
  year={1999},
  volume={14},
  number={5},
  pages={14-23},
  doi={10.1109/5254.796083}}

@inproceedings{horvitz1999principles,
author = {Horvitz, Eric},
title = {Principles of mixed-initiative user interfaces},
year = {1999},
isbn = {0201485591},
publisher = {Association for Computing Machinery},
address = {New York, NY, USA},
url = {https://doi.org/10.1145/302979.303030},
doi = {10.1145/302979.303030},
booktitle = {Proceedings of the SIGCHI Conference on Human Factors in Computing Systems},
pages = {159–166},
numpages = {8},
location = {Pittsburgh, Pennsylvania, USA},
series = {CHI '99}
}

@article{van2021human,
author = {van Berkel, Niels and Skov, Mikael B. and Kjeldskov, Jesper},
title = {Human-AI interaction: intermittent, continuous, and proactive},
year = {2021},
issue_date = {November - December 2021},
publisher = {Association for Computing Machinery},
address = {New York, NY, USA},
volume = {28},
number = {6},
issn = {1072-5520},
url = {https://doi.org/10.1145/3486941},
doi = {10.1145/3486941},
journal = {Interactions},
month = nov,
pages = {67–71},
numpages = {5}
}

@article{vaccaro2024combinations,
  title={When combinations of humans and AI are useful: A systematic review and meta-analysis},
  author={Vaccaro, Michelle and Almaatouq, Abdullah and Malone, Thomas},
  journal={Nature Human Behaviour},
  volume={8},
  number={12},
  pages={2293--2303},
  year={2024},
  publisher={Nature Publishing Group UK London},
  url={https://doi.org/10.1038/s41562-024-02024-1},
  doi={10.1038/s41562-024-02024-1}
}

@inproceedings{cila2022designing,
author = {Cila, Nazli},
title = {Designing Human-Agent Collaborations: Commitment, responsiveness, and support},
year = {2022},
isbn = {9781450391573},
publisher = {Association for Computing Machinery},
address = {New York, NY, USA},
url = {https://doi.org/10.1145/3491102.3517500},
doi = {10.1145/3491102.3517500},
booktitle = {Proceedings of the 2022 CHI Conference on Human Factors in Computing Systems},
articleno = {420},
numpages = {18},
location = {New Orleans, LA, USA},
series = {CHI '22}
}

@inproceedings{castle2025elmo,
author = {Castle-Green, Teresa and Castle-Green, Simon and Lindley, Joseph and Sailaja, Neelima and Lechelt, Susan},
title = {Elmo: An Embodied Conversational Assistant For Community Repair Caf\'{e}s},
year = {2025},
isbn = {9798400715273},
publisher = {Association for Computing Machinery},
address = {New York, NY, USA},
url = {https://doi.org/10.1145/3719160.3737627},
doi = {10.1145/3719160.3737627},
booktitle = {Proceedings of the 7th ACM Conference on Conversational User Interfaces},
articleno = {7},
numpages = {6},
location = {
},
series = {CUI '25}
}

@inproceedings{chen2025need,
author = {Chen, Valerie and Zhu, Alan and Zhao, Sebastian and Mozannar, Hussein and Sontag, David and Talwalkar, Ameet},
title = {Need Help? Designing Proactive AI Assistants for Programming},
year = {2025},
isbn = {9798400713941},
publisher = {Association for Computing Machinery},
address = {New York, NY, USA},
url = {https://doi.org/10.1145/3706598.3714002},
doi = {10.1145/3706598.3714002},
booktitle = {Proceedings of the 2025 CHI Conference on Human Factors in Computing Systems},
articleno = {881},
numpages = {18},
location = {
},
series = {CHI '25}
}

@article{kang2022ai,
  title={AI agency vs. human agency: understanding human--AI interactions on TikTok and their implications for user engagement},
  author={Kang, Hyunjin and Lou, Chen},
  journal={Journal of Computer-Mediated Communication},
  volume={27},
  number={5},
  pages={zmac014},
  year={2022},
  publisher={Oxford University Press},
  url={https://doi.org/10.1093/jcmc/zmac014},
  doi={10.1093/jcmc/zmac014}
}

@article{fanni2023enhancing,
  title={Enhancing human agency through redress in Artificial Intelligence Systems},
  author={Fanni, Rosanna and Steinkogler, Valerie Eveline and Zampedri, Giulia and Pierson, Jo},
  journal={AI \& society},
  volume={38},
  number={2},
  pages={537--547},
  year={2023},
  publisher={Springer},
  url={https://doi.org/10.1007/s00146-022-01454-7},
  doi={10.1007/s00146-022-01454-7}
}

@inproceedings{saleema2019guidelines,
author = {Amershi, Saleema and Weld, Dan and Vorvoreanu, Mihaela and Fourney, Adam and Nushi, Besmira and Collisson, Penny and Suh, Jina and Iqbal, Shamsi and Bennett, Paul N. and Inkpen, Kori and Teevan, Jaime and Kikin-Gil, Ruth and Horvitz, Eric},
title = {Guidelines for Human-AI Interaction},
year = {2019},
isbn = {9781450359702},
publisher = {Association for Computing Machinery},
address = {New York, NY, USA},
url = {https://doi.org/10.1145/3290605.3300233},
doi = {10.1145/3290605.3300233},
booktitle = {Proceedings of the 2019 CHI Conference on Human Factors in Computing Systems},
pages = {1–13},
numpages = {13},
location = {Glasgow, Scotland Uk},
series = {CHI '19}
}

@inproceedings{liu2024compeer,
author = {Liu, Tianjian and Zhao, Hongzheng and Liu, Yuheng and Wang, Xingbo and Peng, Zhenhui},
title = {ComPeer: A Generative Conversational Agent for Proactive Peer Support},
year = {2024},
isbn = {9798400706288},
publisher = {Association for Computing Machinery},
address = {New York, NY, USA},
url = {https://doi.org/10.1145/3654777.3676430},
doi = {10.1145/3654777.3676430},
booktitle = {Proceedings of the 37th Annual ACM Symposium on User Interface Software and Technology},
articleno = {117},
numpages = {22},
location = {Pittsburgh, PA, USA},
series = {UIST '24}
}

@inproceedings{horvitz2013attention,
author = {Horvitz, Eric and Jacobs, Andy and Hovel, David},
title = {Attention-sensitive alerting},
year = {1999},
isbn = {1558606149},
publisher = {Morgan Kaufmann Publishers Inc.},
address = {San Francisco, CA, USA},
booktitle = {Proceedings of the Fifteenth Conference on Uncertainty in Artificial Intelligence},
pages = {305–313},
numpages = {9},
location = {Stockholm, Sweden},
series = {UAI'99}
}

@InProceedings{abowd1999towards,
author={Abowd, Gregory D.
and Dey, Anind K.
and Brown, Peter J.
and Davies, Nigel
and Smith, Mark
and Steggles, Pete},
editor={Gellersen, Hans-W.},
title={Towards a Better Understanding of Context and Context-Awareness},
booktitle={Handheld and Ubiquitous Computing},
doi={10.1007/3-540-48157-5_29},
url={https://doi.org/10.1007/3-540-48157-5_29},
year={1999},
publisher={Springer Berlin Heidelberg},
address={Berlin, Heidelberg},
pages={304--307},
isbn={978-3-540-48157-7}
}

@article{schmidt2000implicit,
  title={Implicit human computer interaction through context},
  author={Schmidt, Albrecht},
  journal={Personal technologies},
  volume={4},
  number={2},
  pages={191--199},
  year={2000},
  publisher={Springer},
  url={https://doi.org/10.1007/BF01324126},
  doi={10.1007/BF01324126}
}

@inproceedings{jameson2002pros,
author = {Jameson, Anthony and Schwarzkopf, Eric},
title = {Pros and Cons of Controllability: An Empirical Study},
year = {2002},
isbn = {3540437371},
publisher = {Springer-Verlag},
address = {Berlin, Heidelberg},
booktitle = {Proceedings of the Second International Conference on Adaptive Hypermedia and Adaptive Web-Based Systems},
pages = {193–202},
numpages = {10},
series = {AH '02}
}

@inproceedings{holter2024deconstructing,
  title={Deconstructing Human-AI Collaboration: Agency, Interaction, and Adaptation},
  author={Holter, Steffen and El-Assady, Mennatallah},
  booktitle={Computer graphics forum},
  volume={43},
  number={3},
  pages={e15107},
  year={2024},
  organization={Wiley Online Library},
  url={https://doi.org/10.1111/cgf.15107},
  doi={10.1111/cgf.15107}
}

@InProceedings{bradshaw2003dimensions,
author={Bradshaw, Jeffrey M.
and Feltovich, Paul J.
and Jung, Hyuckchul
and Kulkarni, Shriniwas
and Taysom, William
and Uszok, Andrzej},
editor={Nickles, Matthias
and Rovatsos, Michael
and Weiss, Gerhard},
title={Dimensions of Adjustable Autonomy and Mixed-Initiative Interaction},
booktitle={Agents and Computational Autonomy},
year={2004},
publisher={Springer Berlin Heidelberg},
address={Berlin, Heidelberg},
pages={17--39},
isbn={978-3-540-25928-2}
}

@ARTICLE{parasuraman2000model,
  author={Parasuraman, R. and Sheridan, T.B. and Wickens, C.D.},
  journal={IEEE Transactions on Systems, Man, and Cybernetics - Part A: Systems and Humans}, 
  title={A model for types and levels of human interaction with automation}, 
  year={2000},
  volume={30},
  number={3},
  pages={286-297},
  doi={10.1109/3468.844354}}

@article{levin2025chatdbg,
author = {Levin, Kyla H. and van Kempen, Nicolas and Berger, Emery D. and Freund, Stephen N.},
title = {ChatDBG: Augmenting Debugging with Large Language Models},
year = {2025},
issue_date = {July 2025},
publisher = {Association for Computing Machinery},
address = {New York, NY, USA},
volume = {2},
number = {FSE},
url = {https://doi.org/10.1145/3729355},
doi = {10.1145/3729355},
journal = {Proc. ACM Softw. Eng.},
month = jun,
articleno = {FSE085},
numpages = {22}
}

@inproceedings{kuang2024enhancing,
author = {Kuang, Emily and Li, Minghao and Fan, Mingming and Shinohara, Kristen},
title = {Enhancing UX Evaluation Through Collaboration with Conversational AI Assistants: Effects of Proactive Dialogue and Timing},
year = {2024},
isbn = {9798400703300},
publisher = {Association for Computing Machinery},
address = {New York, NY, USA},
url = {https://doi.org/10.1145/3613904.3642168},
doi = {10.1145/3613904.3642168},
booktitle = {Proceedings of the 2024 CHI Conference on Human Factors in Computing Systems},
articleno = {3},
numpages = {16},
location = {Honolulu, HI, USA},
series = {CHI '24}
}

@misc{fang2025goldilocks,
      title={The Goldilocks Time Window for Proactive Interventions in Wearable AI Systems}, 
      author={Cathy Mengying Fang and Wazeer Zulfikar and Yasith Samaradivakara and Suranga Nanayakkara and Pattie Maes},
      year={2025},
      eprint={2504.09332},
      archivePrefix={arXiv},
      primaryClass={cs.HC},
      url={https://arxiv.org/abs/2504.09332}, 
}

@article{meurisch2020exploring,
author = {Meurisch, Christian and Mihale-Wilson, Cristina A. and Hawlitschek, Adrian and Giger, Florian and M\"{u}ller, Florian and Hinz, Oliver and M\"{u}hlh\"{a}user, Max},
title = {Exploring User Expectations of Proactive AI Systems},
year = {2020},
issue_date = {December 2020},
publisher = {Association for Computing Machinery},
address = {New York, NY, USA},
volume = {4},
number = {4},
url = {https://doi.org/10.1145/3432193},
doi = {10.1145/3432193},
journal = {Proc. ACM Interact. Mob. Wearable Ubiquitous Technol.},
month = dec,
articleno = {146},
numpages = {22}
}

@article{bradley1958complete,
 ISSN = {01621459, 1537274X},
 URL = {http://www.jstor.org/stable/2281872},
 author = {James V. Bradley},
 journal = {Journal of the American Statistical Association},
 number = {282},
 pages = {525--528},
 publisher = {[American Statistical Association, Taylor & Francis, Ltd.]},
 title = {Complete Counterbalancing of Immediate Sequential Effects in a Latin Square Design},
 urldate = {2026-01-05},
 volume = {53},
 year = {1958}
}

@inproceedings{moore1959shortest,
  title={The shortest path through a maze},
  author={Moore, Edward F},
  booktitle={Proc. of the International Symposium on the Theory of Switching},
  pages={285--292},
  year={1959},
  organization={Harvard University Press}
}

@misc{bodonhelyi2024user,
      title={User Intent Recognition and Satisfaction with Large Language Models: A User Study with ChatGPT}, 
      author={Anna Bodonhelyi and Efe Bozkir and Shuo Yang and Enkelejda Kasneci and Gjergji Kasneci},
      year={2024},
      eprint={2402.02136},
      archivePrefix={arXiv},
      primaryClass={cs.HC},
      url={https://arxiv.org/abs/2402.02136}, 
}

@inproceedings{fossati2010generating,
author = {Fossati, Davide and Di Eugenio, Barbara and Ohlsson, Stellan and Brown, Christopher and Chen, Lin},
title = {Generating proactive feedback to help students stay on track},
year = {2010},
isbn = {364213436X},
publisher = {Springer-Verlag},
address = {Berlin, Heidelberg},
url = {https://doi.org/10.1007/978-3-642-13437-1_56},
doi = {10.1007/978-3-642-13437-1_56},
booktitle = {Proceedings of the 10th International Conference on Intelligent Tutoring Systems - Volume Part II},
pages = {315–317},
numpages = {3},
location = {Pittsburgh, PA},
series = {ITS'10}
}

@article{paul2024smart,
  title={A smart medicine reminder kit with mobile phone calls and some health monitoring features for senior citizens},
  author={Paul, Liton Chandra and Ahmed, Sayed Shifat and Rani, Tithi and Haque, Md Ashraful and Roy, Tushar Kanti and Hossain, Md Najmul and Hossain, Md Azad},
  journal={Heliyon},
  volume={10},
  number={4},
  year={2024},
  publisher={Elsevier},
  url={https://doi.org/10.1016/j.heliyon.2024.e26308},
  doi={10.1016/j.heliyon.2024.e26308}
}

@article{huang2024construction,
  title={Construction and application of medication reminder system: intelligent generation of universal medication schedule},
  author={Huang, Hangxing and Zhang, Lu and Yang, Yongyu and Huang, Ling and Lu, Xikui and Li, Jingyang and Yu, Huimin and Cheng, Shuqiao and Xiao, Jian},
  journal={BioData Mining},
  volume={17},
  number={1},
  pages={23},
  year={2024},
  publisher={Springer},
  url={https://doi.org/10.1186/s13040-024-00376-y},
  doi={10.1186/s13040-024-00376-y}
}

@misc{rooney2009medication,
  title={Medication Reminder System and Method},
  author={Rooney, John and Rooney, Michael},
  year={2009},
  month=feb # "~12",
  publisher={Google Patents},
  note={US Patent App. 12/186,000}
}

@article{parasuraman1997humans,
  title={Humans and automation: Use, misuse, disuse, abuse},
  author={Parasuraman, Raja and Riley, Victor},
  journal={Human factors},
  volume={39},
  number={2},
  pages={230--253},
  year={1997},
  publisher={SAGE Publications Sage CA: Los Angeles, CA},
  url={https://doi.org/10.1518/001872097778543886},
  doi={10.1518/001872097778543886}
}

@inproceedings{nourani2021anchoring,
author = {Nourani, Mahsan and Roy, Chiradeep and Block, Jeremy E and Honeycutt, Donald R and Rahman, Tahrima and Ragan, Eric and Gogate, Vibhav},
title = {Anchoring Bias Affects Mental Model Formation and User Reliance in Explainable AI Systems},
year = {2021},
isbn = {9781450380171},
publisher = {Association for Computing Machinery},
address = {New York, NY, USA},
url = {https://doi.org/10.1145/3397481.3450639},
doi = {10.1145/3397481.3450639},
booktitle = {Proceedings of the 26th International Conference on Intelligent User Interfaces},
pages = {340–350},
numpages = {11},
location = {College Station, TX, USA},
series = {IUI '21}
}

@article{wen2022sense,
  title={The sense of agency in perception, behaviour and human--machine interactions},
  author={Wen, Wen and Imamizu, Hiroshi},
  journal={Nature Reviews Psychology},
  volume={1},
  number={4},
  pages={211--222},
  year={2022},
  publisher={Nature Publishing Group US New York},
  url={https://doi.org/10.1038/s44159-022-00030-6},
  doi={10.1038/s44159-022-00030-6}
}

@article{le2018agency,
  title={Agency modulates interactions with automation technologies},
  author={Le Goff, Kevin and Rey, Arnaud and Haggard, Patrick and Oullier, Olivier and Berberian, Bruno},
  journal={Ergonomics},
  volume={61},
  number={9},
  pages={1282--1297},
  year={2018},
  publisher={Taylor \& Francis},
  url={https://doi.org/10.1080/00140139.2018.1468493},
  doi={10.1080/00140139.2018.1468493}
}

@article{berberian2012automation,
    doi = {10.1371/journal.pone.0034075},
    author = {Berberian, Bruno AND Sarrazin, Jean-Christophe AND Le Blaye, Patrick AND Haggard, Patrick},
    journal = {PLOS ONE},
    publisher = {Public Library of Science},
    title = {Automation Technology and Sense of Control: A Window on Human Agency},
    year = {2012},
    month = {03},
    volume = {7},
    url = {https://doi.org/10.1371/journal.pone.0034075},
    pages = {1-6},
    number = {3},
}

@article{vereschak2021evaluate,
author = {Vereschak, Oleksandra and Bailly, Gilles and Caramiaux, Baptiste},
title = {How to Evaluate Trust in AI-Assisted Decision Making? A Survey of Empirical Methodologies},
year = {2021},
issue_date = {October 2021},
publisher = {Association for Computing Machinery},
address = {New York, NY, USA},
volume = {5},
number = {CSCW2},
url = {https://doi.org/10.1145/3476068},
doi = {10.1145/3476068},
journal = {Proc. ACM Hum.-Comput. Interact.},
month = oct,
articleno = {327},
numpages = {39}
}

@article{vantrepotte2022leveraging,
  title={Leveraging human agency to improve confidence and acceptability in human-machine interactions},
  author={Vantrepotte, Quentin and Berberian, Bruno and Pagliari, Marine and Chambon, Val{\'e}rian},
  journal={Cognition},
  volume={222},
  pages={105020},
  year={2022},
  publisher={Elsevier},
  url={https://doi.org/10.1016/j.cognition.2022.105020},
  doi={10.1016/j.cognition.2022.105020}
}

@article{koedinger2007exploring,
  title={Exploring the assistance dilemma in experiments with cognitive tutors},
  author={Koedinger, Kenneth R and Aleven, Vincent},
  journal={Educational psychology review},
  volume={19},
  number={3},
  pages={239--264},
  year={2007},
  publisher={Springer},
  url={https://doi.org/10.1007/s10648-007-9049-0},
  doi={10.1007/s10648-007-9049-0}
}

@article{newman2022helping,
  title={Helping people through space and time: Assistance as a perspective on human-robot interaction},
  author={Newman, Benjamin A and Aronson, Reuben M and Kitani, Kris and Admoni, Henny},
  journal={Frontiers in Robotics and AI},
  volume={8},
  pages={720319},
  year={2022},
  publisher={Frontiers Media SA},
  url={https://doi.org/10.3389/frobt.2021.720319},
  doi={10.3389/frobt.2021.720319}
}

@article{nair2025solicit,
  title={To Solicit or Not to Solicit? Impact of AI Assistance Delivery Mechanisms on Decision-Making},
  author={Nair, Pavithra PM and Gressel, Gilad and Anand, Malavika Nambrath and Achuthan, Krishnashree},
  journal={International Journal of Human--Computer Interaction},
  pages={1--24},
  year={2025},
  publisher={Taylor \& Francis},
  url={https://doi.org/10.1080/10447318.2025.2536617},
  doi={10.1080/10447318.2025.2536617}
}

@book{nunnally1994psychometric,
  title={Psychometric Theory 3E},
  author={Nunnally, J.C.},
  isbn={9780071070881},
  lccn={93022756},
  series={McGraw-Hill series in psychology},
  year={1994},
  publisher={Tata McGraw-Hill Education}
}

@article{cronbach1951coefficient,
  title={Coefficient alpha and the internal structure of tests},
  author={Cronbach, Lee J},
  journal={psychometrika},
  volume={16},
  number={3},
  pages={297--334},
  year={1951},
  publisher={Springer-Verlag},
  url={https://doi.org/10.1007/BF02310555},
  doi={10.1007/BF02310555}
}

@article{brooks2017glmmtmb,
  title={glmmTMB balances speed and flexibility among packages for zero-inflated generalized linear mixed modeling},
  author={Brooks, Mollie E and Kristensen, Kasper and Van Benthem, Koen J and Magnusson, Arni and Berg, Casper W and Nielsen, Anders and Skaug, Hans J and M{\"a}chler, Martin and Bolker, Benjamin M},
  year={2017},
  journal = {The R Journal},
  doi = {10.32614/RJ-2017-066},
  pages = {378--400},
  volume = {9},
  number = {2},
}

@article{hartig2016dharma,
  title={DHARMa: residual diagnostics for hierarchical (multi-level/mixed) regression models},
  author={Hartig, Florian},
  journal={CRAN: Contributed Packages},
  year={2016},
  publisher={The R Foundation}
}

@article{kubinec2023ordered,
  title={Ordered beta regression: a parsimonious, well-fitting model for continuous data with lower and upper bounds},
  author={Kubinec, Robert},
  journal={Political analysis},
  volume={31},
  number={4},
  pages={519--536},
  year={2023},
  url={https://doi.org/10.1017/pan.2022.20},
  doi={10.1017/pan.2022.20}
}

@article{shono2008application,
  title={Application of the Tweedie distribution to zero-catch data in CPUE analysis},
  author={Shono, Hiroshi},
  journal={Fisheries Research},
  volume={93},
  number={1-2},
  pages={154--162},
  year={2008},
  publisher={Elsevier},
  url={https://doi.org/10.1016/j.fishres.2008.03.006},
  doi={10.1016/j.fishres.2008.03.006}
}

@book{aho2013foundational,
  title={Foundational and applied statistics for biologists using R},
  author={Aho, Ken A},
  year={2013},
  publisher={CRC Press},
  url={https://doi.org/10.1201/b16126},
  isbn={9780429062988}
}

@book{fox2018r,
  title={An R companion to applied regression},
  author={Fox, John and Weisberg, Sanford},
  year = {2019},
  publisher = {Sage},
  address = {Thousand Oaks {CA}},
  url = {https://www.john-fox.ca/Companion/},
}

@article{gale2013method,
  title={Using the framework method for the analysis of qualitative data in multi-disciplinary health research},
  author={Gale, Nicola K and Heath, Gemma and Cameron, Elaine and Rashid, Sabina and Redwood, Sabi},
  journal={BMC medical research methodology},
  volume={13},
  number={1},
  pages={117},
  year={2013},
  publisher={Springer},
  url={https://doi.org/10.1186/1471-2288-13-117},
  doi={10.1186/1471-2288-13-117}
}

@inproceedings{nikolaidis2017human,
author = {Nikolaidis, Stefanos and Zhu, Yu Xiang and Hsu, David and Srinivasa, Siddhartha},
title = {Human-Robot Mutual Adaptation in Shared Autonomy},
year = {2017},
isbn = {9781450343367},
publisher = {Association for Computing Machinery},
address = {New York, NY, USA},
url = {https://doi.org/10.1145/2909824.3020252},
doi = {10.1145/2909824.3020252},
booktitle = {Proceedings of the 2017 ACM/IEEE International Conference on Human-Robot Interaction},
pages = {294–302},
numpages = {9},
location = {Vienna, Austria},
series = {HRI '17}
}

@incollection{schaefer2016measuring,
  title={Measuring trust in human robot interactions: Development of the “trust perception scale-HRI”},
  author={Schaefer, Kristin E},
  booktitle={Robust intelligence and trust in autonomous systems},
  pages={191--218},
  year={2016},
  publisher={Springer},
  url={https://doi.org/10.1007/978-1-4899-7668-0_10},
  doi={10.1007/978-1-4899-7668-0_10}
}

@incollection{hart1988development,
title = {Development of NASA-TLX (Task Load Index): Results of Empirical and Theoretical Research},
editor = {Peter A. Hancock and Najmedin Meshkati},
series = {Advances in Psychology},
publisher = {North-Holland},
volume = {52},
pages = {139-183},
year = {1988},
booktitle = {Human Mental Workload},
issn = {0166-4115},
doi = {https://doi.org/10.1016/S0166-4115(08)62386-9},
url = {https://www.sciencedirect.com/science/article/pii/S0166411508623869},
author = {Sandra G. Hart and Lowell E. Staveland},
}

@article{franke2019personal,
  title={A personal resource for technology interaction: development and validation of the affinity for technology interaction (ATI) scale},
  author={Franke, Thomas and Attig, Christiane and Wessel, Daniel},
  journal={International Journal of Human--Computer Interaction},
  volume={35},
  number={6},
  pages={456--467},
  year={2019},
  publisher={Taylor \& Francis},
  url={https://doi.org/10.1080/10447318.2018.1456150},
  doi={10.1080/10447318.2018.1456150}
}

@article{davis1989perceived,
 ISSN = {02767783, 21629730},
 URL = {http://www.jstor.org/stable/249008},
 author = {Fred D. Davis},
 journal = {MIS Quarterly},
 number = {3},
 pages = {319--340},
 publisher = {Management Information Systems Research Center, University of Minnesota},
 title = {Perceived Usefulness, Perceived Ease of Use, and User Acceptance of Information Technology},
 urldate = {2026-01-05},
 volume = {13},
 year = {1989}
}

@article{hoegl2001teamwork,
  title={Teamwork quality and the success of innovative projects: A theoretical concept and empirical evidence},
  author={Hoegl, Martin and Gemuenden, Hans Georg},
  journal={Organization science},
  volume={12},
  number={4},
  pages={435--449},
  year={2001},
  publisher={Informs},
  url={https://doi.org/10.1287/orsc.12.4.435.10635},
  doi={10.1287/orsc.12.4.435.10635}
}

@article{simon1955behavioral,
  title={A behavioral model of rational choice},
  author={Simon, Herbert A},
  journal={The quarterly journal of economics},
  pages={99--118},
  year={1955},
  publisher={JSTOR},
  url={https://doi.org/10.2307/1884852},
  doi={10.2307/1884852}
}

@misc{nargund2025understanding,
      title={Understanding Mode Switching in Human-AI Collaboration: Behavioral Insights and Predictive Modeling}, 
      author={Avinash Ajit Nargund and Arthur Caetano and Kevin Yang and Rose Yiwei Liu and Philip Tezaur and Kriteen Shrestha and Qisen Pan and Tobias Höllerer and Misha Sra},
      year={2025},
      eprint={2509.20666},
      archivePrefix={arXiv},
      primaryClass={cs.HC},
      url={https://arxiv.org/abs/2509.20666}, 
}

@inproceedings{bixler2015automatic,
author = {Bixler, Robert and Blanchard, Nathaniel and Garrison, Luke and D'Mello, Sidney},
title = {Automatic Detection of Mind Wandering During Reading Using Gaze and Physiology},
year = {2015},
isbn = {9781450339124},
publisher = {Association for Computing Machinery},
address = {New York, NY, USA},
url = {https://doi.org/10.1145/2818346.2820742},
doi = {10.1145/2818346.2820742},
booktitle = {Proceedings of the 2015 ACM on International Conference on Multimodal Interaction},
pages = {299–306},
numpages = {8},
location = {Seattle, Washington, USA},
series = {ICMI '15}
}

@article{steyvers2024three,
  title={Three challenges for AI-assisted decision-making},
  author={Steyvers, Mark and Kumar, Aakriti},
  journal={Perspectives on Psychological Science},
  volume={19},
  number={5},
  pages={722--734},
  year={2024},
  publisher={Sage Publications Sage CA: Los Angeles, CA},
  url={https://doi.org/10.1177/17456916231181102},
  doi={10.1177/17456916231181102}
}

@article{wu2025human,
  title={Human-generative AI collaboration enhances task performance but undermines human’s intrinsic motivation},
  author={Wu, Suqing and Liu, Yukun and Ruan, Mengqi and Chen, Siyu and Xie, Xiao-Yun},
  journal={Scientific Reports},
  volume={15},
  number={1},
  pages={15105},
  year={2025},
  publisher={Nature Publishing Group UK London},
  url={https://doi.org/10.1038/s41598-025-98385-2},
  doi={10.1038/s41598-025-98385-2}
}

@article{pataranutaporn2021ai,
  title={AI-generated characters for supporting personalized learning and well-being},
  author={Pataranutaporn, Pat and Danry, Valdemar and Leong, Joanne and Punpongsanon, Parinya and Novy, Dan and Maes, Pattie and Sra, Misha},
  journal={Nature Machine Intelligence},
  volume={3},
  number={12},
  pages={1013--1022},
  year={2021},
  publisher={Nature Publishing Group UK London},
  url={https://doi.org/10.1038/s42256-021-00417-9},
  doi={10.1038/s42256-021-00417-9}
}

@misc{pataranutaporn2025my,
      title={"My Boyfriend is AI": A Computational Analysis of Human-AI Companionship in Reddit's AI Community}, 
      author={Pat Pataranutaporn and Sheer Karny and Chayapatr Archiwaranguprok and Constanze Albrecht and Auren R. Liu and Pattie Maes},
      year={2025},
      eprint={2509.11391},
      archivePrefix={arXiv},
      primaryClass={cs.HC},
      url={https://arxiv.org/abs/2509.11391}, 
}

@article{endsley1995out,
  title={The out-of-the-loop performance problem and level of control in automation},
  author={Endsley, Mica R and Kiris, Esin O},
  journal={Human factors},
  volume={37},
  number={2},
  pages={381--394},
  year={1995},
  publisher={SAGE Publications Sage CA: Los Angeles, CA},
  doi = {10.1518/001872095779064555},
  url={https://doi.org/10.1518/001872095779064555},
  doi = {10.1518/001872095779064555}
}

\appendix

\section{Post-Study Likert-Scale Questions}
\label{post-study-questionnaire}

\subsection{AI's Competence and Reliability}

\begin{itemize}
    \item The AI was competent.
    \item The AI understood the puzzle well.
    \item The AI performed reliably.
\end{itemize}

\subsection{AI's Helpfulness and Effectiveness}
\begin{itemize}
    \item The AI helped me solve the puzzle more effectively.
    \item The AI made helpful contributions in solving the puzzle.
    \item The AI cared about helping me solve the puzzle.
    \item Overall, the AI was helpful.
\end{itemize}

\subsection{Quality and Sufficiency of AI Offered Help}

\begin{itemize}
    \item The AI provided all the moves I needed.
    \item The AI helped me at the right time.
    \item The AI's help was not enough.
    \item I was able to ask the AI for the right amount of help.
\end{itemize}

\subsection{AI's Interpretability}

\begin{itemize}
    \item I understood why the AI made its moves.
    \item It felt like the AI suggestions were aligned with what I was thinking.
    \item The AI's help derailed my thinking.
\end{itemize}

\subsection{Teamwork with AI}
\begin{itemize}
    \item It felt like AI and I were working as a team.
    \item The AI and I had a shared goal.
    \item I think I solved the puzzle or me + AI solved the puzzle together.
    \item I was satisfied with the way the AI and I worked together.
    \item There is good fit between what the AI offered and what I needed from it.
\end{itemize}

\section{Post-Study Open-Ended Questions}
\label{post-study-open-ended}

\subsection{Questions for AI-Assisted Conditions (Button and Timer)}

\begin{itemize}
    \item What did you like about the AI's behavior in the puzzle task?
    \item What did you dislike about the AI's behavior in the puzzle task?
    \item When solving the puzzles, what strategies (if any) did you use to more effectively solve the puzzle?
    \item What should this AI do differently to become your ideal puzzle solving assistant?
    \item Did the number of moves AI helped with impact how you solved the puzzle?
    \item How did you determine how many moves of help you needed?
\end{itemize}

\subsection{Questions for the Control Condition}

In this section, you will provide suggestions for designing an AI assistant to support users in tasks and problem-solving scenarios, such as Rush Hour.

\begin{itemize}
    \item When solving the puzzles, what strategies (if any) did you use to more effectively solve the puzzle?
    \item At any point, did you feel stuck? If so, what would have helped you get unstuck?
    \item What kind of support would you find most helpful from the AI? (e.g., hints, written explanations, visual demonstrations, step-by-step guidance, etc.)
    \item Would you like the AI to explain why it suggests a certain move, or just show the move? Why?
    \item At what moments would you prefer the AI to assist you — continuously, upon request, after multiple failed attempts, or under other conditions?
    \item Would you prefer the AI to act like a coach, a teammate, or a tool? Why?
    \item Should the AI ask you questions to help you reflect, or just offer suggestions?
    \item How much control do you want to retain while solving the puzzle?
    \item Would you like feedback after solving a puzzle, such as alternative shorter solutions or strategy tips?
\end{itemize}

\end{document}